\documentclass[article]{jss}

\usepackage{graphicx,graphics,subfigure}
\usepackage{multicol,multirow}

\author{Blake MacDonald\\Acadia University \And
        Pritam Ranjan\\Acadia University \And
        Hugh Chipman \\ Acadia University}
\title{\pkg{GPfit}: An \proglang{R} package for Gaussian Process Model Fitting using a New Optimization Algorithm}

\Plainauthor{Blake MacDonald, Pritam Ranjan, Hugh Chipman} 
\Plaintitle{GPfit: An R package for Gaussian Process Model Fitting using a New Optimization Algorithm} 
\Shorttitle{\pkg{GPfit}: An \proglang{R} package for GP model fitting} 

\Abstract{
Gaussian process (GP) models are commonly used statistical metamodels
for emulating expensive computer simulators. Fitting a GP model can be
numerically unstable if any pair of design points in the input space are
close together. \cite{ranjanNugget} proposed a computationally stable
approach for fitting GP models to deterministic computer simulators. They
used a genetic algorithm based approach that is robust but computationally
intensive for maximizing the likelihood. This paper implements a
slightly modified version of the model proposed by \cite{ranjanNugget}, as
the new \proglang{R} package \pkg{GPfit}.  A novel parameterization of the spatial correlation function and a new multi-start gradient based optimization
algorithm yield optimization that is robust and typically faster than the
genetic algorithm based approach. We present two examples with \proglang{R} codes
to illustrate the usage of the main functions in \pkg{GPfit}. Several test
functions are used for performance comparison with a popular \proglang{R} package
\pkg{mlegp}. \pkg{GPfit} is a free software and distributed under the general
public license, as part of the \proglang{R} software project \citep{R_software}.
}
\Plainkeywords{Computer experiments, clustering, near-singularity, nugget} 

\Address{
  Pritam Ranjan\\
  Department of Mathematics and Statistics\\
  Acadia University\\
  15 University Avenue, Wolfville, NS, Canada\\
  E-mail: \email{pritam.ranjan@acadiau.ca}\\
  URL: \url{http://acadiau.ca/~pranjan/}
}

\begin{document}


\section{Introduction} \label{sec:intro}

Computer simulators are often used to model complex physical and
engineering processes that are either infeasible, too expensive or time
consuming to observe. Examples include tracking the population
for bowhead whales in Western Arctic \citep{poole}, monitoring traffic
control system \citep{medina}, and dynamics of dark energy and
dark matter in cosmological studies \citep{arbey}. Realistic computer
simulators can still be computationally expensive to run, and they are
often approximated (or emulated) using statistical models. \cite{gp1}
proposed emulating such an expensive deterministic simulator as a
realization of a Gaussian stochastic process (GP). This paper presents
a new \proglang{R} package \pkg{GPfit} for robust and computationally
efficient fitting of GP models to deterministic
simulator outputs.

The computational stability of GP estimation algorithms can depend
critically on the set of design points and corresponding simulator outputs
that are used to build a GP model. If any pair of design points in the input
space are close together, the spatial correlation matrix $R$ may become
near-singular and hence the GP model fitting procedure computationally
unstable. A popular approach to overcome this numerical instability
is to introduce a small ``nugget" parameter $\delta$ in the model,
i.e., $R$ is replaced by $R_\delta = R + \delta I$, that is estimated
along with the other model parameters (e.g., \cite{neal, booker,
santner2003daa, gramacy}). However, adding a nugget in the model
introduces additional smoothing in the predictor and as a result the
predictor is no longer an  interpolator. Thus, it is challenging to
choose an appropriate value of $\delta$ that maintains the delicate
balance between the stabilization and minimizing the over-smoothing of
the model predictions. \citet{ranjanNugget} proposed a computationally
stable approach by introducing a lower bound on the nugget, which minimizes
unnecessary over-smoothing and improves the model accuracy.

Instead of trying to interpolate the data, one may argue that all
simulators are noisy and the statistical surrogates should always smooth
the simulator data (e.g., \cite{gramacy_nugget}). In spite of the recent
interest in stochastic simulators (e.g., \cite{poole}, \cite{arbey}),
deterministic simulators are still being actively used. For instance,
\cite{medina} demonstrate the preference of deterministic traffic
simulators over their stochastic counterparts. The model considered in
\pkg{GPfit} assumes that the computer simulator is deterministic and
is very similar to the GP model proposed in \cite{ranjanNugget}.

The maximum likelihood approach for fitting the GP model requires
optimizing the log-likelihood, which can often have multiple local optima
\citep{Yuan,Schirru,lawrence,petelin}. This makes the model fitting
procedure computationally challenging. \cite{ranjanNugget} uses a genetic
algorithm (GA) approach, which is robust but computationally intensive for
likelihood optimization. \pkg{GPfit} uses a multi-start gradient based
search algorithm that is robust and typically faster than the GA used in
\cite{ranjanNugget}. A clustering based approach on a large space-filling
design over the parameter space is used for choosing the
initial values of the gradient search. Furthermore, we proposed a new
parameterization of the spatial correlation function for the ease of
likelihood optimization.

The remainder of the paper is organized as
follows. Section~\ref{sec:method} presents a brief review of
the GP model in \cite{ranjanNugget}, the new parameterization
of the correlation function and the new optimization algorithm
implemented in \pkg{GPfit}. In Section~\ref{sec:gpfit}, the main
functions of \pkg{GPfit} and their arguments are discussed. Two
examples illustrating the usage of \pkg{GPfit} are presented in
Section~\ref{sec:illus_examples}. Section~\ref{sec:mlegp_comp} compares
\pkg{GPfit} with other popular \proglang{R} packages. This
includes an empirical performance comparison with the popular \proglang{R}
package \pkg{mlegp}. The paper concludes with a few remarks
in Section~\ref{sec:conclusion}.

\section{Methodology}\label{sec:method}
Section~\ref{sec:GPmodel} reviews the GP model proposed
in \cite{ranjanNugget} (for more details on GP models, see
\cite{santner2003daa} and \cite{rasmussen2006book}). We propose a
new parameterization of the correlation function in
Section~\ref{sec:reparameterization} that facilitates
optimization of the likelihood. The new optimization
algorithm implemented in \pkg{GPfit} is presented in
Section~\ref{sec:optimization}.

\subsection{Gaussian process model}\label{sec:GPmodel}
Let the $i$-th input and the corresponding output of the computer simulator
be denoted by a $d$-dimensional vector, $x_i = (x_{i1}, ..., x_{id})'$ and
$y_{i} = y(x_i)$ respectively. The experimental design $D_{0} = \{x_1, ...,
x_n\}$ is the set of $n$ input trials stored in an $n \times d$ matrix $X$.
We assume  $x_i \in [0, 1]^d$. The outputs are held in the $n \times 1$ vector $Y = y(X) = (y_1, \dots , y_n)'$. The simulator output, $y(x_i)$, is modeled as
$$
y(x_i) = \mu + z(x_i);\hspace{20pt} i = 1, ..., n,
$$
where $\mu$ is the overall mean, and $z(x_i)$ is a GP with $E(z(x_i))
= 0$, $Var(z(x_i)) = \sigma^2$, and $Cov(z(x_i), z(x_j)) = \sigma^2
R_{ij}$. In general, $y(X)$ has a multivariate normal distribution,
$N_n(\mathbf{1_n}\mu,\Sigma)$, where $\Sigma = \sigma^2R$ is formed
with correlation matrix $R$ having elements $R_{ij}$, and
$\mathbf{1_n}$ is a $n \times 1$ vector of all ones. Although there
are several choices for the correlation structure, we follow
\cite{ranjanNugget} and use the Gaussian correlation function given by
\begin{equation}\label{eqn:corr_gp}
R_{ij} = \prod_{k=1}^d \exp\{-\theta_k|x_{ik} - x_{jk}|^{2}\},\qquad \textrm{for all} \quad i, j,
\end{equation}
where $\theta = (\theta_1, ..., \theta_d) \in [0, \infty)^d$ is a vector of hyper-parameters. The closed form estimators of $\mu$ and $\sigma^2$ given by
$$
   \hat{\mu}(\theta) = {({\bf1_n}'R^{-1}{\bf1_n})}^{-1}({\bf1_n}'R^{-1}Y)
   \ \textrm{and} \   \hat{\sigma}^2(\theta) =
   \frac{(Y-{\bf1_n}\hat{\mu}(\theta))'R^{-1}(Y-{\bf1_n}\hat{\mu}(\theta))}{n},
$$
are used to obtain the negative profile log-likelihood (hereonwards, referred to as deviance)
$$
  -2\log (L_{\theta}) \propto \log(|R|) + n \log[(Y-{\bf1_n}\hat{\mu}(\theta))' R^{-1}(Y-{\bf1_n}\hat{\mu}(\theta))],
$$
for estimating the hyper-parameters $\theta$, where $|R|$ denotes the determinant of $R$.

Following the maximum likelihood approach, the best linear unbiased predictor at $x^*$ (as shown in \cite{gp1}) is
$$
\hat{y}(x^*) = \hat{\mu} + r'R^{-1}(Y - \mathbf{1_n}\hat{\mu}) = \left[\frac{(1-r'R^{-1}\mathbf{1_n})}{\mathbf{1_n}'R^{-1}\mathbf{1_n}}\mathbf{1_n}' +r'\right] R^{-1}Y = C'Y,
$$
with mean squared error
\begin{eqnarray*}
s^2(x^*) &=& E\left[(\hat{y}(x^*) - y(x^*))^2\right] \\
&=&  \sigma^2 (1 - 2C'r + C'RC) =  \sigma^2 \left( 1 - r'R^{-1}r + \frac{(1-\mathbf{1_n}'R^{-1}r)^2}{\mathbf{1_n}R^{-1}\mathbf{1_n}}\right),
\end{eqnarray*}
where $r = (r_1(x^*), ..., r_n(x^*)), \textrm{and } r_i(x^*) = corr(z(x^*), z(x_i))$. In practice, the parameters $\mu$, $\sigma^2$ and $\theta$ are replaced with their respective estimates.

Fitting a GP model to $n$ data points requires the repeated computation of the determinant and inverse of the $n \times n$ correlation matrix $R$. Such correlation matrices are positive definite by definition, however, the computation of $|R|$ and $R^{-1}$ can sometimes be unstable due to near-singularity. An $n \times n$ matrix $R$ is said to be near-singular (or, ill-conditioned) if its condition number $\kappa(R) = \|R\|\cdot\|R^{-1}\|$ is too large, where $\|\cdot\|$ denotes the $L_2$--matrix norm (see \cite{ranjanNugget} for details). Near-singularity prohibits precise computation of the deviance and hence the parameter estimates. This is a common problem in fitting GP models which occurs if any pair of design points in the input space are close together \citep{neal}. A popular approach to overcome near-singularity is to introduce a small nugget or jitter parameter, $\delta \in (0,1)$, in the model (i.e., $R$ is replaced by $R_\delta = R + \delta I$) that is estimated along with the other model parameters.

Replacing $R$ with $R_{\delta}$ in the GP model introduces additional
smoothing of the simulator data that is undesirable for emulating a
deterministic simulator. \citet{ranjanNugget} proposed a computationally
stable approach to choosing the nugget parameter $\delta$.
They introduced a lower bound on $\delta$ that
minimizes the unnecessary over-smoothing. The lower bound given by
\citet{ranjanNugget} is
\begin{equation}\label{eqn:deltalb}
\delta_{lb} = \max \left\{ \frac{\lambda_n(\kappa(R)-e^a)}{\kappa(R)(e^a-1)}, 0 \right\},
\end{equation}
where $\lambda_n$ is the largest eigenvalue of $R$ and $e^a$
is the threshold of $\kappa(R)$ that ensures a well conditioned
$R$. \cite{ranjanNugget} suggest $a = 25$ for space-filling Latin
hypercube designs (LHDs) \citep{mckay}.

\pkg{GPfit} uses the GP model with $R_{\delta_{lb}} = R + \delta_{lb} I$. The \proglang{R} package \pkg{mlegp}, used for
performance comparison of \pkg{GPfit} in Section~\ref{sec:mlegp_comp},
implements the classical GP model with $R$ replaced by $R_{\delta} = R +
\delta I$, and estimates $\delta$ along with other hyper-parameters by
minimizing the deviance. In both approaches the deviance function happens
to be bumpy with multiple local optima. Next, we investigate a
novel parameterization of the correlation function that makes the deviance easier to optimize.

\subsection{Reparameterization of the correlation
function}\label{sec:reparameterization}
\begin{figure}[htb!]\centering
\subfigure[~]{\label{fig:old_eg1d_log_lik} \includegraphics[width=3in]{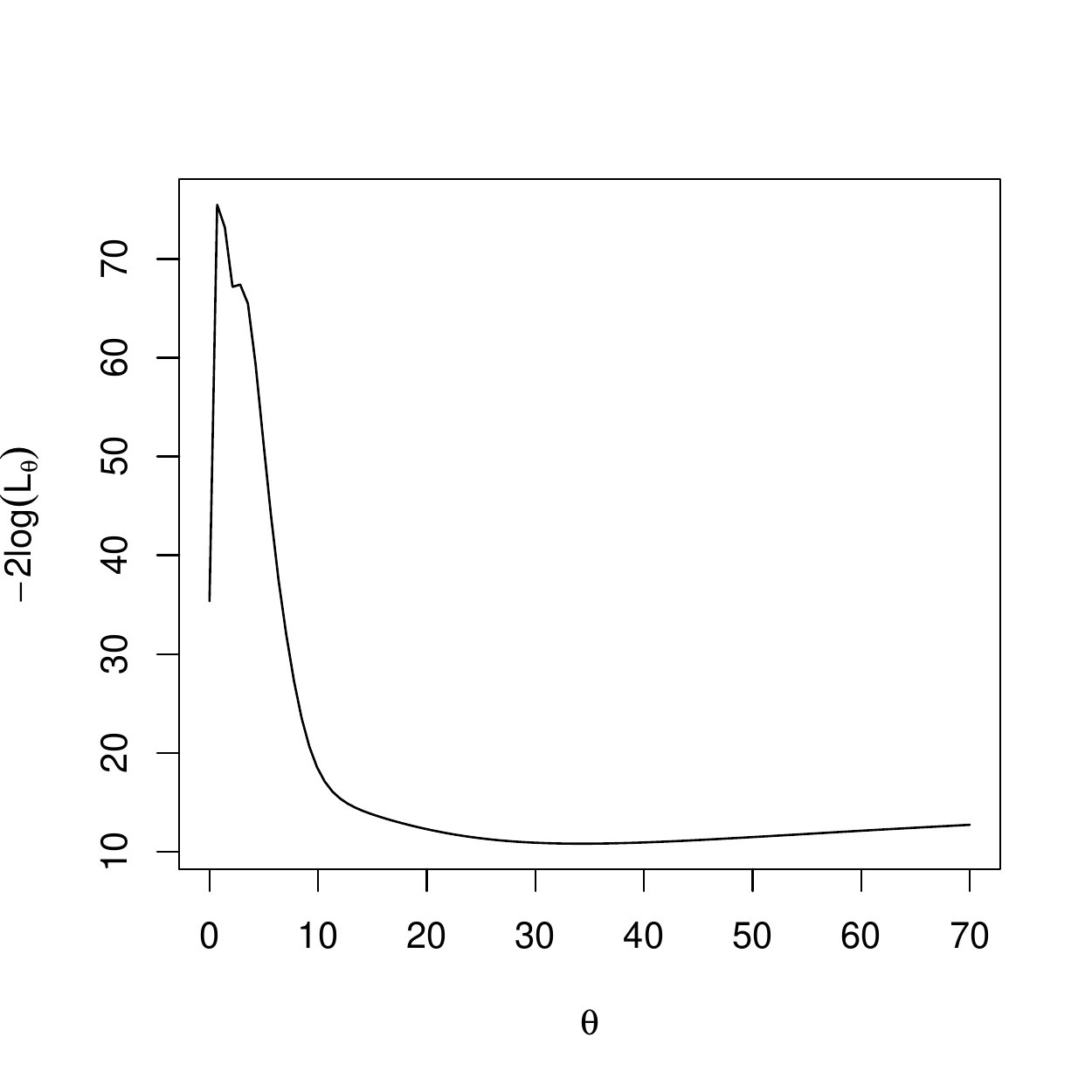}}\hspace{-0.25cm}
\subfigure[~]{\label{fig:old_eg1d_log_lik} \includegraphics[width=3in]{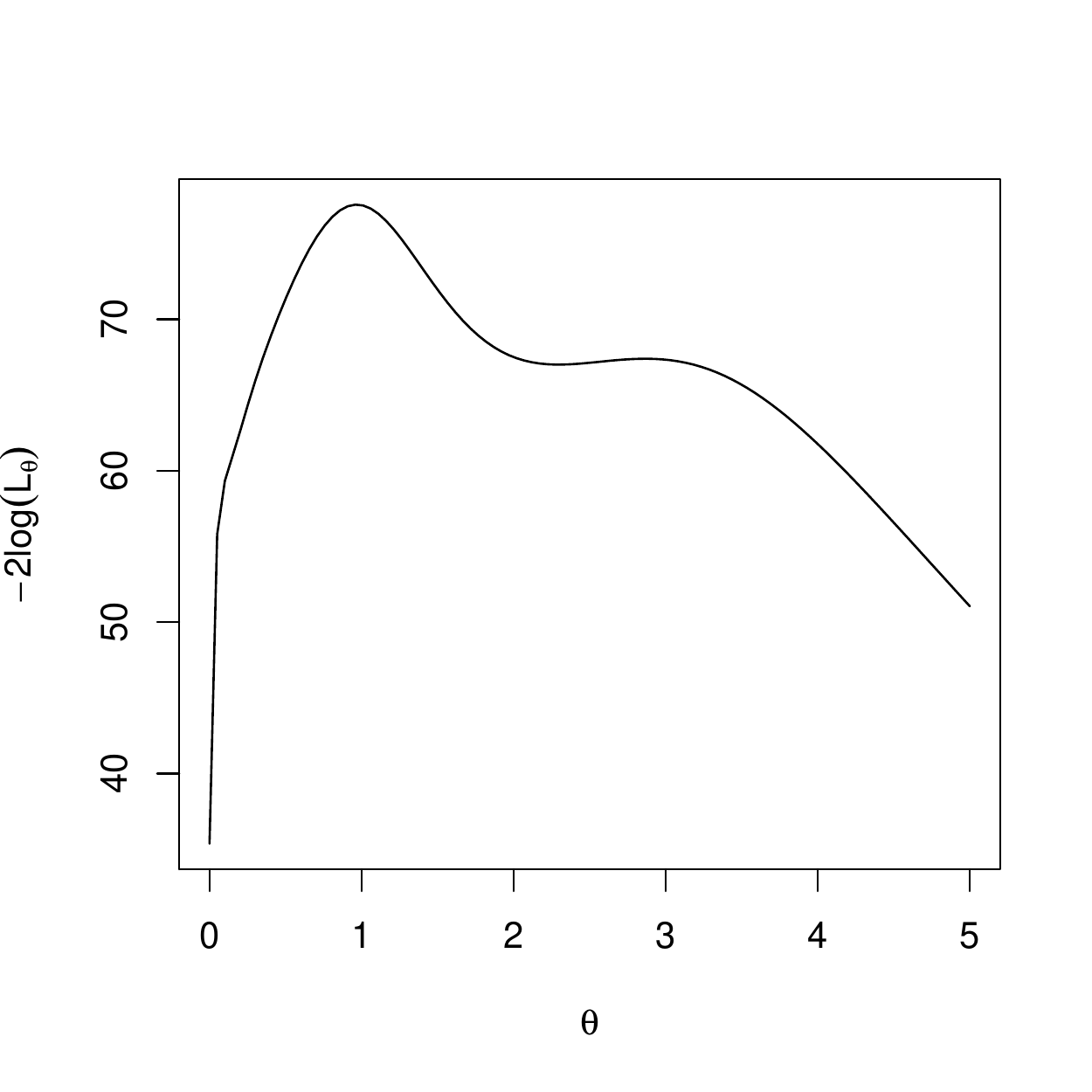}}
\subfigure[~]{\label{fig:old_eg2d_log_lig} \includegraphics[width=3in]{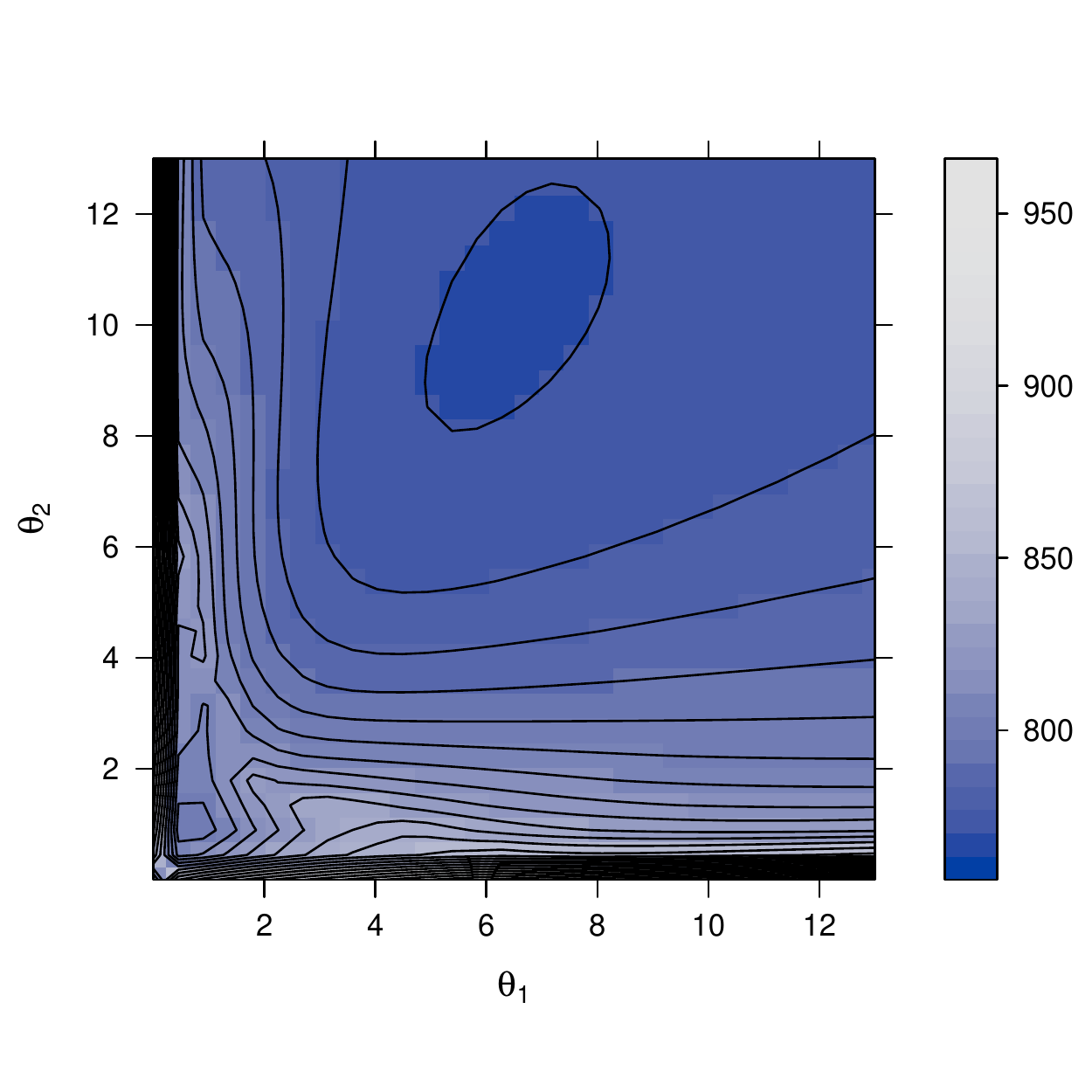}}\hspace{-0.25cm}
\subfigure[~]{\label{fig:old_eg2d_log_lig} \includegraphics[width=3in]{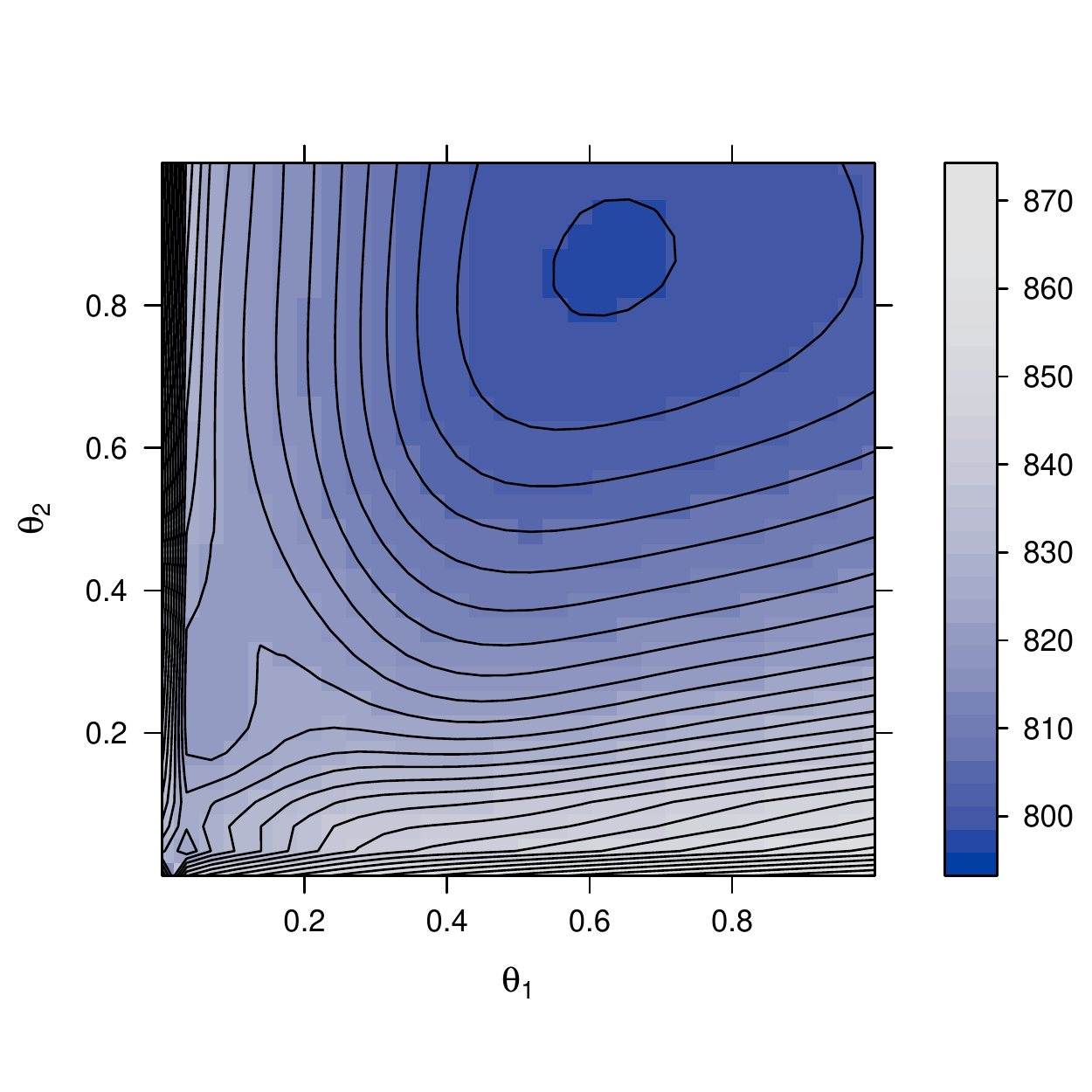}}
\caption{The plots show deviance (\ref{eqn:lik_theta}) w.r.t. the GP parameter(s)
$\theta$. Panels (a) and (b) correspond to Example~1 (with
$d=1, n=10$), and (c) and (d) display deviance for Example~2 (with $d=2, n=30$).
Panels (b) and (d) are enlargements of (a) and (b) near 0, respectively.
}\label{fig:old_eg_log_lik}\end{figure}

The key component of fitting the GP model described in Section~\ref{sec:GPmodel} is the estimation of the correlation parameters by minimizing the deviance
\begin{equation}\label{eqn:lik_theta}
  -2\log (L_{\theta}) \propto \log(|R_{\delta_{lb}}|) + n \log[(Y-{\bf1_n}\hat{\mu}(\theta))' R^{-1}_{\delta_{lb}}(Y-{\bf1_n}\hat{\mu}(\theta))].
\end{equation}
The deviance surface can be bumpy and have several local
optima. For instance, the deviance functions for two examples
in Section~\ref{sec:illus_examples} are displayed in
Figure~\ref{fig:old_eg_log_lik}.

Figure~\ref{fig:old_eg_log_lik} shows that the deviance function is bumpy
near $\theta=0$ and there are multiple local optima. Evolutionary
algorithms like GA (used by \cite{ranjanNugget}) are often robust for
such objective functions, however, they can be computationally intensive
(especially, because the computational cost of $|R|$ and $R^{-1}$ is
$O(n^3)$ and evolutionary algorithms often employ many evaluations
of the objective function). Gradient-based optimization
might be faster but will require careful selection of initial
values to achieve the global minimum of the deviance function.  It may be
tempting to use a space-filling design over the parameter space for the
stating points, however, such designs (e.g., maximin LHD) often tend to
stay away from the boundaries and corners. This is unfavourable because
the deviance functions (Figure~\ref{fig:old_eg_log_lik}) are very active
near $\theta=0$.

\begin{figure}[htb!]\centering
\subfigure[~]{\label{fig:new_eg1d_log_lik} \includegraphics[width=3in]{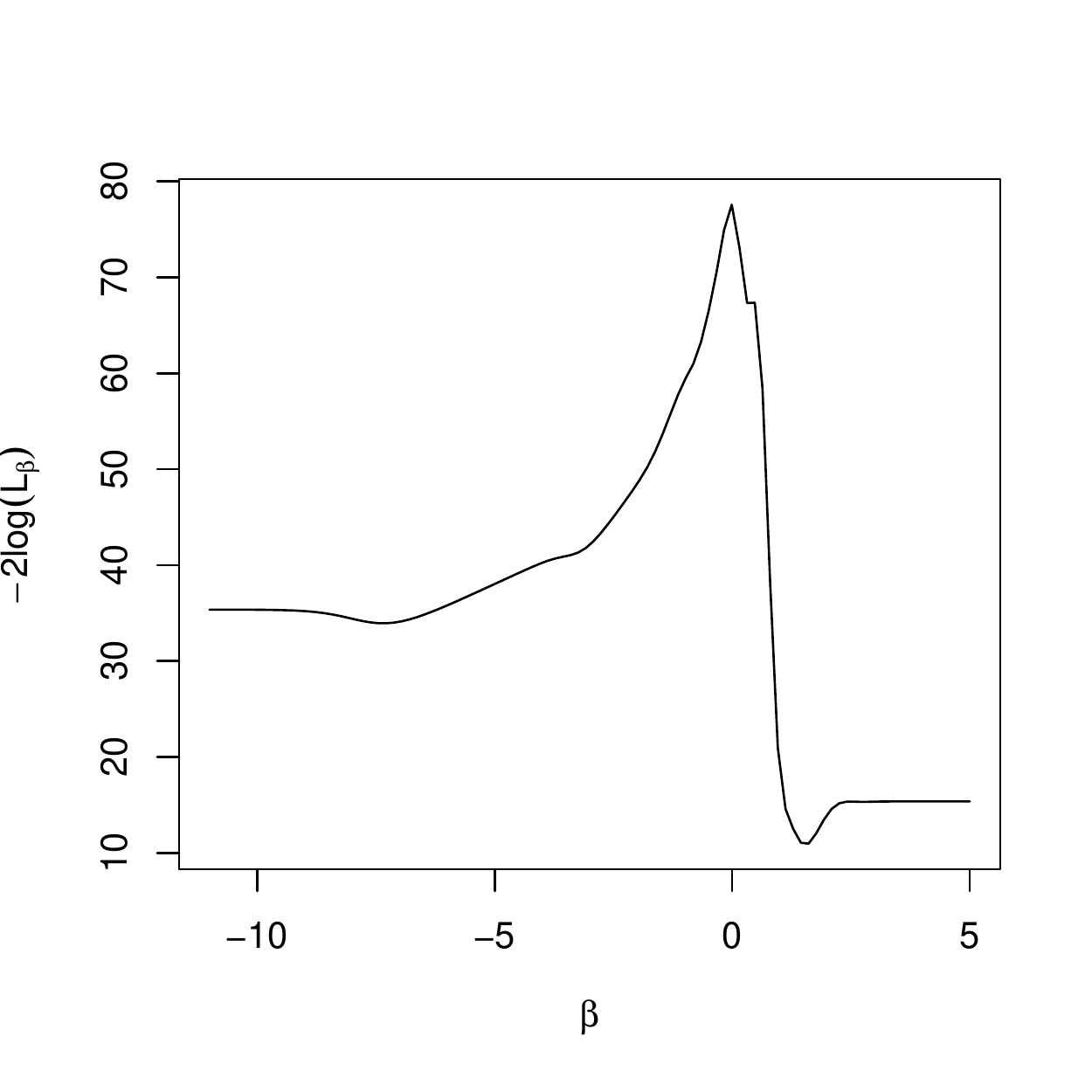}}\hspace{-0.25cm}
\subfigure[~]{\label{fig:new_eg1d_log_lik} \includegraphics[width=3in]{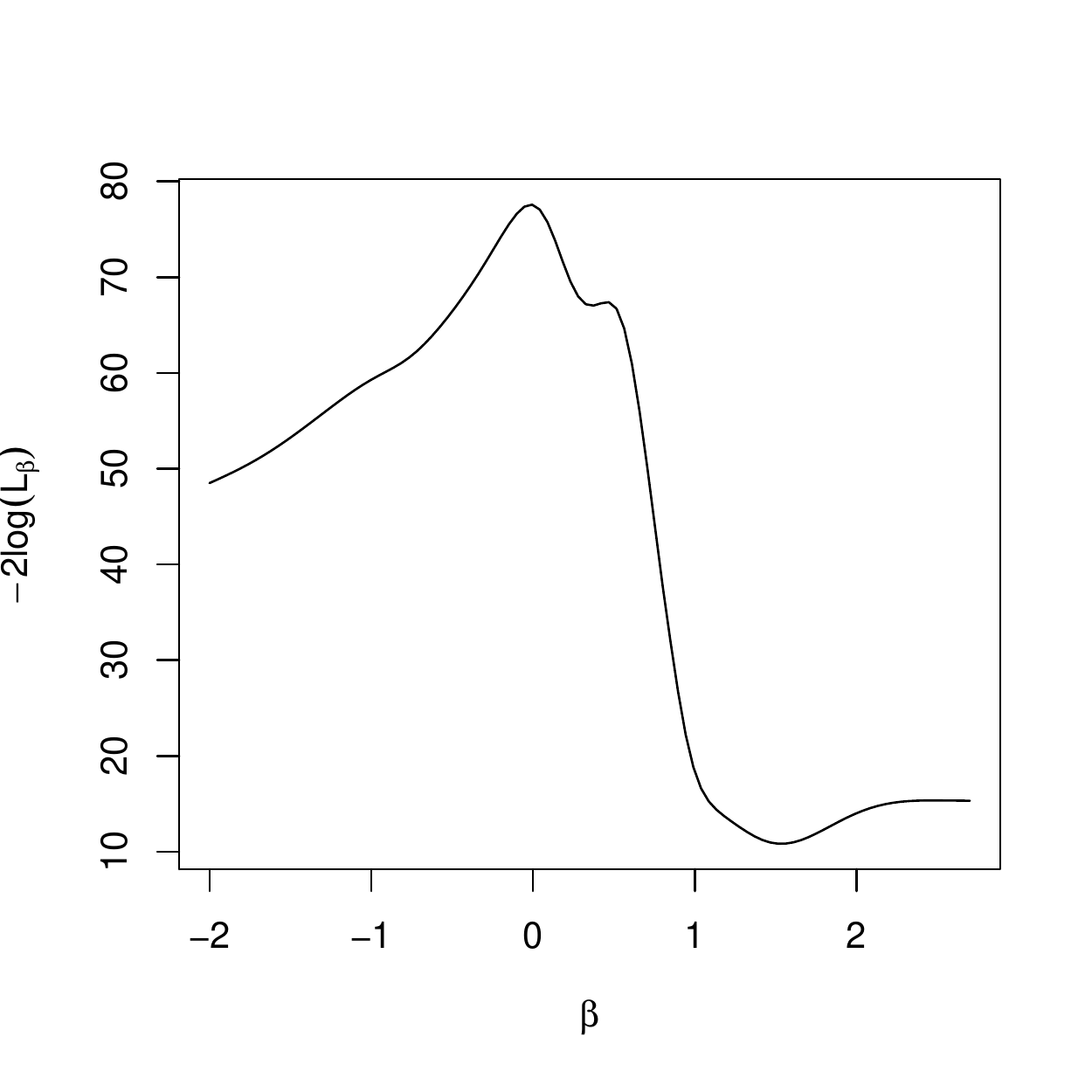}}
\subfigure[~]{\label{fig:new_eg2d_log_lig} \includegraphics[width=3in]{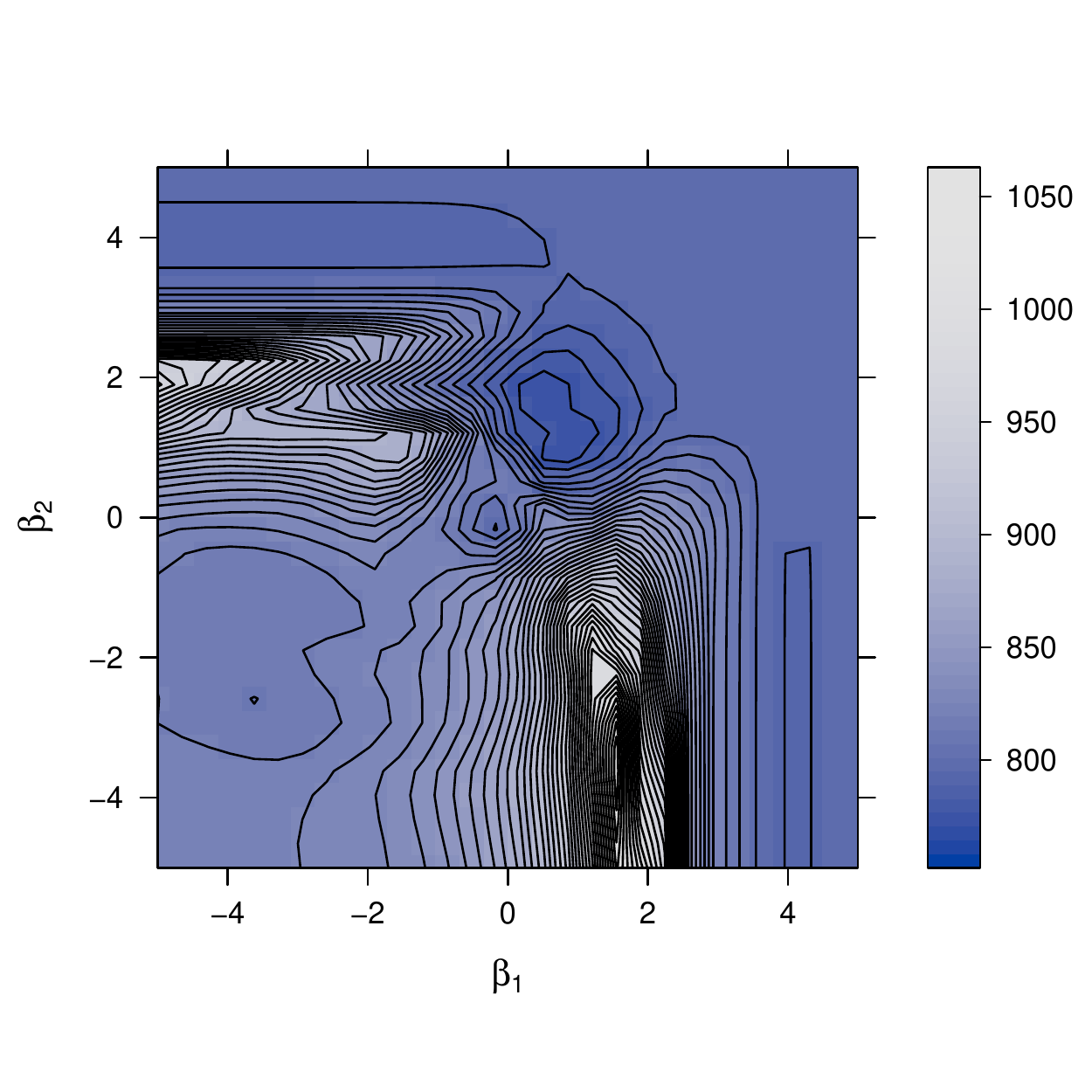}}\hspace{-0.25cm}
\subfigure[~]{\label{fig:new_eg2d_log_lig} \includegraphics[width=3in]{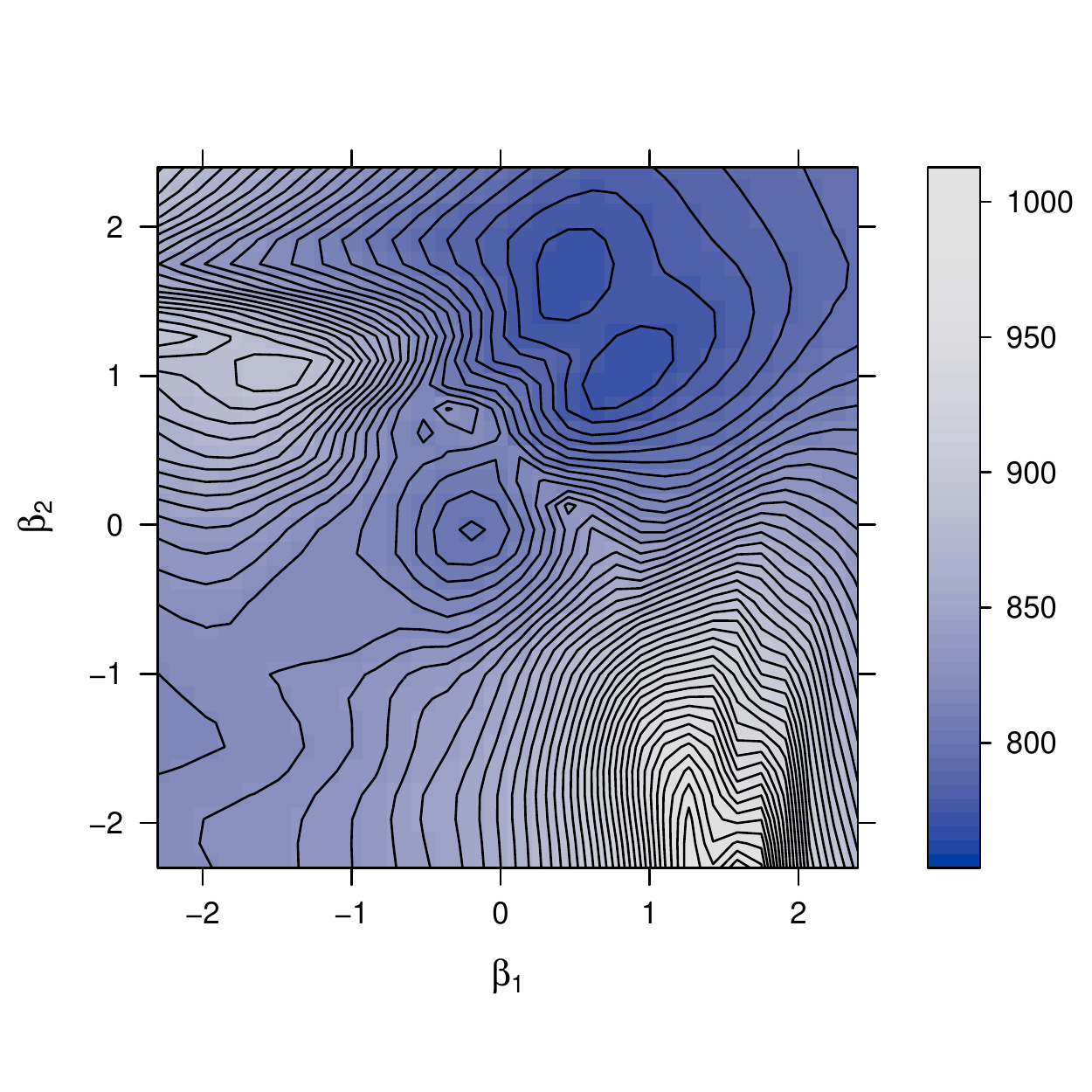}}
\caption{The plots show deviance under $\beta$ parameterization of $R$
(\ref{eqn:corr_beta}), for the same examples and data as in
Figure~\ref{fig:old_eg_log_lik}.
Panels (a) and (b) correspond to Example~1 (with
$d=1, n=10$), and (c) and (d) display deviance for Example~2 (with $d=2, n=30$).
Panels (b) and (d) are enlargements of (a) and (b) near 0, respectively.
}\label{fig:new_eg_log_lik}\end{figure}

To address the issue of a bumpy deviance surface near the
boundaries of the parameter space, we propose a new parameterization of $R$. Let $\beta_k = \log_{10}(\theta_k)$ for $k=1, ..., d$, then
\begin{equation}\label{eqn:corr_beta}
R_{ij} =  \prod_{k=1}^{d}\exp\left\{-10^{\beta_k}|x_{ik}-x_{jk}|^2\right\}, \qquad \textrm{for all} \quad i, j,
\end{equation}
where a small value of $\beta_k$ implies a very high spatial correlation or
a relatively flat surface in the $k$-th coordinate, and the large values of
$\beta_k$ imply low correlation, or a very wiggly surface with respect to
the $k$-th input factor. Figure~\ref{fig:new_eg_log_lik} displays the two
deviance surfaces (shown in Figure~\ref{fig:old_eg_log_lik}) under the
$\beta$ - parameterization of $R$ (\ref{eqn:corr_beta}). Though the new
parameterization of $R$ (\ref{eqn:corr_beta}) results in an unbounded
parameter space $\Omega = (-\infty, \infty)^d$, the peaks and
dips of the deviance surface are now in the middle of the search space.
This should facilitate a thorough search through the local
optima and the choice of a set of initial values for a gradient based search.
%
%

\pkg{GPfit} uses a multi-start gradient based search algorithm for
minimizing the deviance. The gradient based approach is often
computationally fast, and careful selection of the multiple initial
values of the search algorithm makes our implementation robust.

\subsection{Optimization algorithm}\label{sec:optimization}
A standard gradient based search algorithm like L-BFGS-B \citep{byrd} finds
the local optimum closest to the initial value, and thus often gets stuck in the wrong local optima. Our objective is to find $\beta$ that minimizes the deviance function. \cite{lawrence} argue that a slightly suboptimal solution of the deviance optimization problem may not always be a threat in the GP model setup, as alternative interpretations can be used to justify the model fit. However, the prediction accuracy at unsampled locations may suffer from suboptimal parameter estimates.
In an attempt to obtain a good fit of the GP model, \pkg{GPfit} uses a
multi-start L-BFGS-B algorithm for optimizing the deviance $-2 \log
(L_{\beta})$. We first find a subregion $\Omega_0$ of the parameter space
$\Omega=(-\infty, \infty)^d$ that is likely to contain the optimal
parameter values. Then, a set of initial values for L-BFGS-B is carefully chosen to cover $\Omega_0$.

The structural form of the spatial correlation function (\ref{eqn:corr_beta}) guarantees that its value lies in $[0, 1]$. That is, excluding the extreme cases of perfectly correlated and absolutely uncorrelated observations, $R_{ij}$ can be approximately bounded as:
$$
\exp\{-5\} = 0.0067 \le R_{ij} \le 0.9999 = \exp\{-10^{-4}\},
$$
or equivalently,
$$
10^{-4} \leq \sum_{k=1}^{d}10^{\beta_k}|x_{ik}-x_{jk}|^2 \leq 5.
$$
To convert the bounds above into workable ranges for the $\beta_k$,
we need to consider ranges for $|x_{ik}-x_{jk}|$.  Assuming the objective is to approximate the overall simulator surface in $[0, 1]^d$, \cite{loeppky} argue that $n=10\cdot d$ is a good rule of thumb for determining the size of a space-filling design over the input locations of the simulator. In this case, the maximum value of the minimum inter-point distance along $k$-th coordinate is $|x_{ik}-x_{jk}| \approx 1/10$. Furthermore, if we also make a simplifying assumption that the simulator is equally smooth in all directions, i.e., $\beta_k = \beta_0$, then the inequality simplifies to
\begin{equation}\label{eqn:beta_range}
-2 - \log_{10}(d) \le \beta_k \le \log_{10}(500) - \log_{10}(d).
\end{equation}

That is, $\Omega_0 = \{(\beta_1, ..., \beta_d)\ :\ -2 - \log_{10}(d) \le
\beta_k \le \log_{10}(500) - \log_{10}(d), k=1, ..., d \}$ is the set of
$\beta=(\beta_1, ..., \beta_d)$ values that is likely to contain the likelihood optimizer. We use $\Omega_0$ for restricting the initial values of L-BFGS-B algorithm to a manageable area, and the optimal solutions can be found outside this range.

The initial values for L-BFGS-B can be chosen using a large
space-filling LHD on $\Omega_0$. However, Figure~\ref{fig:new_eg_log_lik}
shows that some parts of the likelihood surface are roughly flat, and
multiple starts of L-BFGS-B in such regions might be unnecessary.
We use a combination of k-means clustering applied to the design of
parameter values, and evaluation of the deviance to reduce a large LHD to a
more manageable set of initial values. Since the construction of $\Omega_0$ assumed the simplification $\beta_k = \beta_0$ for all $k$, and in some cases, for instance, in
Figure~\ref{fig:new_eg2d_log_lig}, the deviance surface appears symmetric in the
two coordinates, we enforce the inclusion of an additional initial
value of L-BFGS-B on the main diagonal of $\Omega_0$. This diagonal point is the best of three L-BFGS-B runs only along the main diagonal, $\beta_k = \beta_0$ for all $k$.

The deviance optimization algorithm is summarized as follows:
\begin{enumerate}
\item{Choose a 200$d$-point maximin LHD for $\beta=(\beta_1, ..., \beta_d)$ in the hyper-rectangle $\Omega_0$.}
\item{Choose the $80d$ values of $\beta$ that correspond to the smallest $-2\log(L_{\beta})$ values.}
\item{Use k-means clustering algorithm on these $80d$ points to find $2d$ groups. To improve the quality of the clusters, five random restarts of k-means are used.}
\item{For $d \geq 2$, run L-BFGS-B algorithm along the main diagonal of $\Omega_0$ starting at three equidistant points on the diagonal (i.e., at 25\%, 50\% and 75\%). Choose the best of the three L-BFGS-B outputs, i.e., with smallest $-2\log(L_{\beta})$ value.}
\item{These $2d+1$ (or 2 if $d = 1$) initial values, found in Steps 3 and 4, are then used in the L-BFGS-B routine to find the smallest $-2\log(L_{\beta})$ and corresponding $\hat{\beta}_{mle} \in \Omega$. }
\end{enumerate}

The multi-start L-BFGS-B algorithm outlined above requires $\left(200d +
\sum_{i=1}^{2d+1}\eta_i + \sum_{j=1}^{3}\eta_j' \right)$ deviance
evaluations, where $\eta_i$ is the number of deviance evaluations for the
$i$-th L-BFGS-B run in $\Omega$ space, and $\eta_j'$ is the number of
deviance evaluations for the $j$-th L-BFGS-B run along the diagonal of the
$\Omega_0$ space. For every iteration of L-BFGS-B, the algorithm computes
one gradient (i.e., $2d$ deviance evaluations) and adaptively finds the
location of the next step. That is, $\eta_i$ and $\eta_j'$ may vary, and
the total number of deviance evaluations in the optimization process cannot
be determined. Nonetheless, the empirical evidence based on the examples in
Sections~\ref{sec:illus_examples} and \ref{sec:mlegp_comp} suggest that the
optimization algorithm used here is much faster than the GA in
\cite{ranjanNugget} which uses $1000 d^2$ evaluations of
(\ref{eqn:lik_theta}) for fitting the GP model in $d$-dimensional input
space. Both deviance minimization approaches have a few tunable parameters,
for instance, the initial values and the maximum number of iterations (\code{maxit}) in L-BFGS-B, and the population size and number of generations in a GA, that can perhaps be adjusted to get better performance (i.e., fewer deviance calls to achieve the same accuracy in optimizing the deviance surface).

\section{GPfit package}\label{sec:gpfit}
In this section, we discuss different functions of \pkg{GPfit} that
implements our proposed model, which is the computationally stable version
of the GP model proposed by \cite{ranjanNugget} with the new
parameterization of correlation matrix $R$ (Section~\ref{sec:reparameterization}), and optimization algorithm described in Section~\ref{sec:optimization}.

The main functions for the users of \pkg{GPfit} are \code{GP_fit()},
\code{predict()} and (for $d \le 2$) \code{plot()}. Both \code{predict()} and \code{plot()} use \code{GP_fit()}class objects for providing prediction and plots respectively. The code for fitting the GP model to $n$ data points in $d$-dimensional input space
stored in an $n\times d$ matrix \code{X} and an $n-$ vector
\code{Y} is:
\begin{CodeChunk}
\begin{CodeInput}
    GP_fit(X, Y, control=c(200*d,80*d,2*d), nug_thres=20,
                                            trace=FALSE, maxit=100)
\end{CodeInput}
\end{CodeChunk}

The default values of \code{`control'}, \code{`nug_thres'}, \verb"`trace'" and \verb"`maxit'" worked smoothly for all the examples implemented in this paper, however, they can be changed if necessary.
\begin{itemize}
\item \verb"control:" A vector of three tunable parameters used in the deviance optimization algorithm. The default values correspond to choosing \verb"2*d" clusters (using k-means clustering algorithm) based on \verb"80*d" best points (smallest deviance) from a \verb"200*d" - point random maximin LHD in $\Omega_0$.

\item \verb"nug_thres:" A threshold parameter used in the calculation of the lower bound of the nugget, $\delta_{lb}$. Although \cite{ranjanNugget} suggest \verb"nug_thres=25" for space-filling designs, we use a conservative default value  \verb"nug_thres=20". This value might change for different design schemes.

\item \verb"trace:" A flag that indicates whether or not to print the information on the final runs of the L-BFGS-B algorithm. The default \verb"trace=FALSE" implies no printing.

\item \code{maxit:} is the maximum number of iterations per L-BFGS-B run in the deviance optimization. We use the \verb"optim" package default \verb"`maxit=100'".
\end{itemize}

\verb"GP_fit()" returns the object of class \verb"GP" that contains the
data set \verb"X", \verb"Y" and the estimated model parameters
$\hat{\beta}, \hat{\sigma}^2$ and $\delta_{lb}(\hat{\beta})$. Assuming
\verb"GPmodel" is the \verb"GP" class object, \verb"print(GPmodel,...)" presents the values of the object \verb"GPmodel", and options like \verb"digits" can be used for ``...". As an alternative, one can use \verb"summary(GPmodel)" to get the same output.

If \verb"xnew" contains the set of unobserved inputs, \verb"`predict(GPmodel, xnew)'" returns the predicted response $\hat{y}(x^*)$ and the associated MSE $s^2(x^*)$ for every input $x^*$ in \verb"xnew". It also returns a data frame with the predictions combined with the \verb"xnew". The expressions of $\hat{y}(x^*)$ and $s^2(x^*)$ are shown in Section~\ref{sec:GPmodel} subject to the replacement of $R$ with $R_{\delta_{lb}(\hat{\beta}_{mle})} = R + \delta_{lb}(\hat{\beta}_{mle}) I$. The default value of \verb"xnew" is the design matrix \verb"X" used for model fitting.

The plotting function \verb"plot()" takes the \verb"GP" object as input and depicts the model predictions and the associated MSEs over a regular grid of the $d$-dimensional input space
for $d = 1$ and $2$. Various graphical options can be specified as additional
arguments:
\begin{CodeChunk}
\begin{CodeInput}
    plot(GPmodel, range=c(0, 1), resolution=50, colors=c('black',
    'blue', 'red'), line_type=c(1, 1), pch=1, cex=2, surf_check=FALSE,
    response=TRUE, ...)
\end{CodeInput}
\end{CodeChunk}

For $d=1$, \verb"plot()" generates the predicted response $\hat{y}(x)$ and uncertainty bounds $\hat{y}(x) \pm 2 s(x)$ over a regular grid of \verb"`resolution'" many points in the specified \verb"range=c(0, 1)". The graphical arguments \verb"colors", \verb"line_type", \verb"pch" and \verb"cex" are only applicable for one-dimensional plots. One can also provide additional graphical argument in ``..." for changing the plots (see \verb"`par'" in the base \proglang{R} function \verb"`plot()'").

For $d=2$, the default arguments of \verb"plot()" with \verb"GP" object produces a level plot of $\hat{y}(x^*)$. The plots are based on the model prediction using \verb"predict()" at a \verb"resolution" $\times$ \verb"resolution" regular grid over $[0, 1]^2$. The argument \verb"surf_check=TRUE" can be used to generate a surface plot instead, and MSEs can be plotted by using \verb"response=FALSE". Options like \verb"shade" and \verb"drape" from \code{wireframe()} function, \verb"contour" and \verb"cuts" from \code{levelplot()} function in \pkg{lattice} \citep{lattice}, and color specific arguments in \pkg{colorspace} \citep{seq_hcl,colorspace} can also be passed in for ``...".

\section{Examples using GPfit}\label{sec:illus_examples}
This section demonstrates the usage of \pkg{GPfit} functions and
the interpretation of the outputs of the main functions.
Two test functions are used as computer simulators
to illustrate the functions of this package.\\

\textbf{Example 1} Let $x \in [0, 1]$, and the computer simulator output, $y(x)$, be generated using the simple one-dimensional test function
$$ y(x) = \log(x+0.1) + \sin(5\pi x),$$
referred to as the function \verb"computer_simulator" below. Suppose we wish to fit the GP model to a data set collected over a random maximin LHD of size $n=7$. The design can be generated using the \verb"maximinLHS" function in the \proglang{R} package \pkg{lhs} \citep{lhs,stein1987}. The following \proglang{R} code shows how to load the packages, generate the simulator outputs and then fit the GP model using \verb"GP_fit()".
\begin{CodeChunk}
\begin{CodeInput}
R> library("GPfit")
R> library("lhs")
R> n = 7
R> x = maximinLHS(n,1)
R> y = matrix(0,n,1)
R> for(i in 1:n){ y[i] = computer_simulator(x[i]) }
R> GPmodel = GP_fit(x,y)
\end{CodeInput}
\end{CodeChunk}
The proposed optimization algorithm used only 227 deviance evaluations for
fitting this GP model. The parameter estimates of the fitted GP model are
obtained using \verb"print(GPmodel)". For printing only four significant
decimal places, \verb"digits=4" can be used in \verb"print()".
\begin{CodeChunk}
\begin{CodeOutput}
Number Of Observations: n = 7
Input Dimensions: d = 1

Correlation: Exponential (power = 2)
Correlation Parameters:
    beta_hat
[1]    1.977

sigma^2_hat: [1] 0.7444
delta_lb(beta_hat): [1] 0
nugget threshold parameter: 20
\end{CodeOutput}
\end{CodeChunk}

\begin{figure}[htb!]\centering
\includegraphics[width=4.0in]{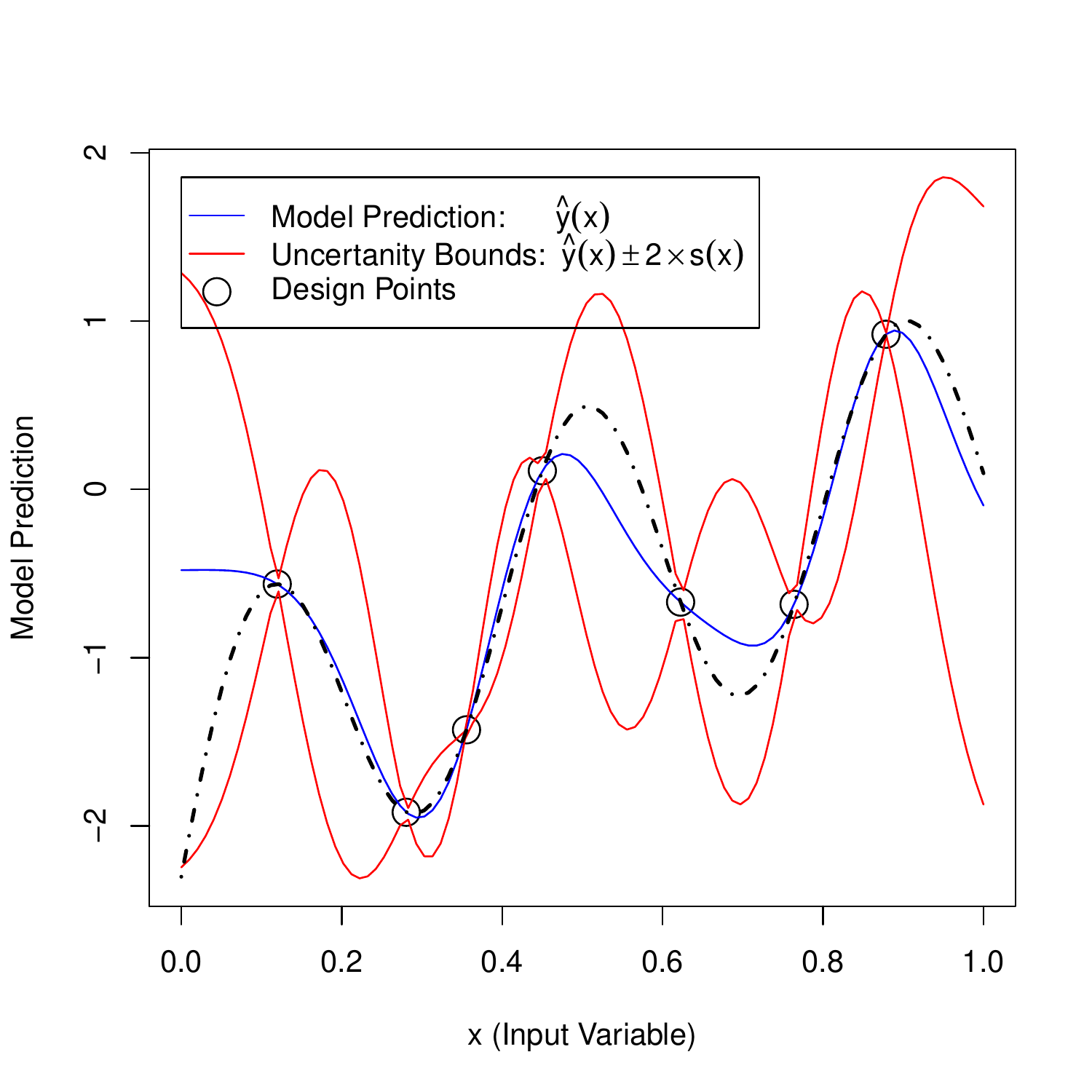}\vspace{-0.75cm}
\caption{The plot shows the model predictions and uncertainty bands for Example~1. The true simulator output curve is also displayed by the dash-dotted line. }\label{fig:eg1_0}
\end{figure}

\begin{figure}[htb!]\centering
\subfigure[Default arguments]{\label{fig:eg1a} \includegraphics[width=3in]{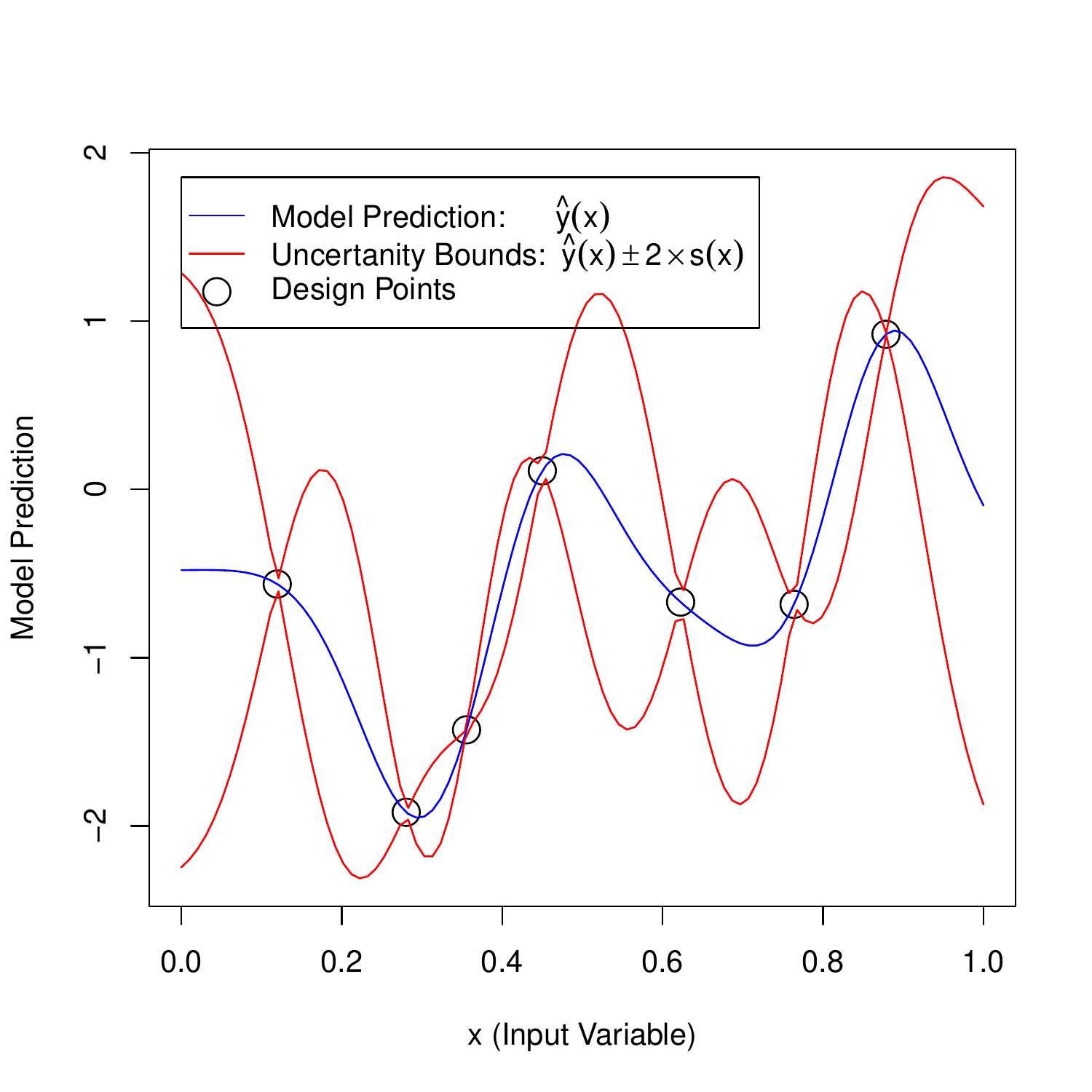}}\hspace{-0.75cm}
\subfigure[\code{line\_type=c(1,2)}]{\label{fig:eg1b} \includegraphics[width=3in]{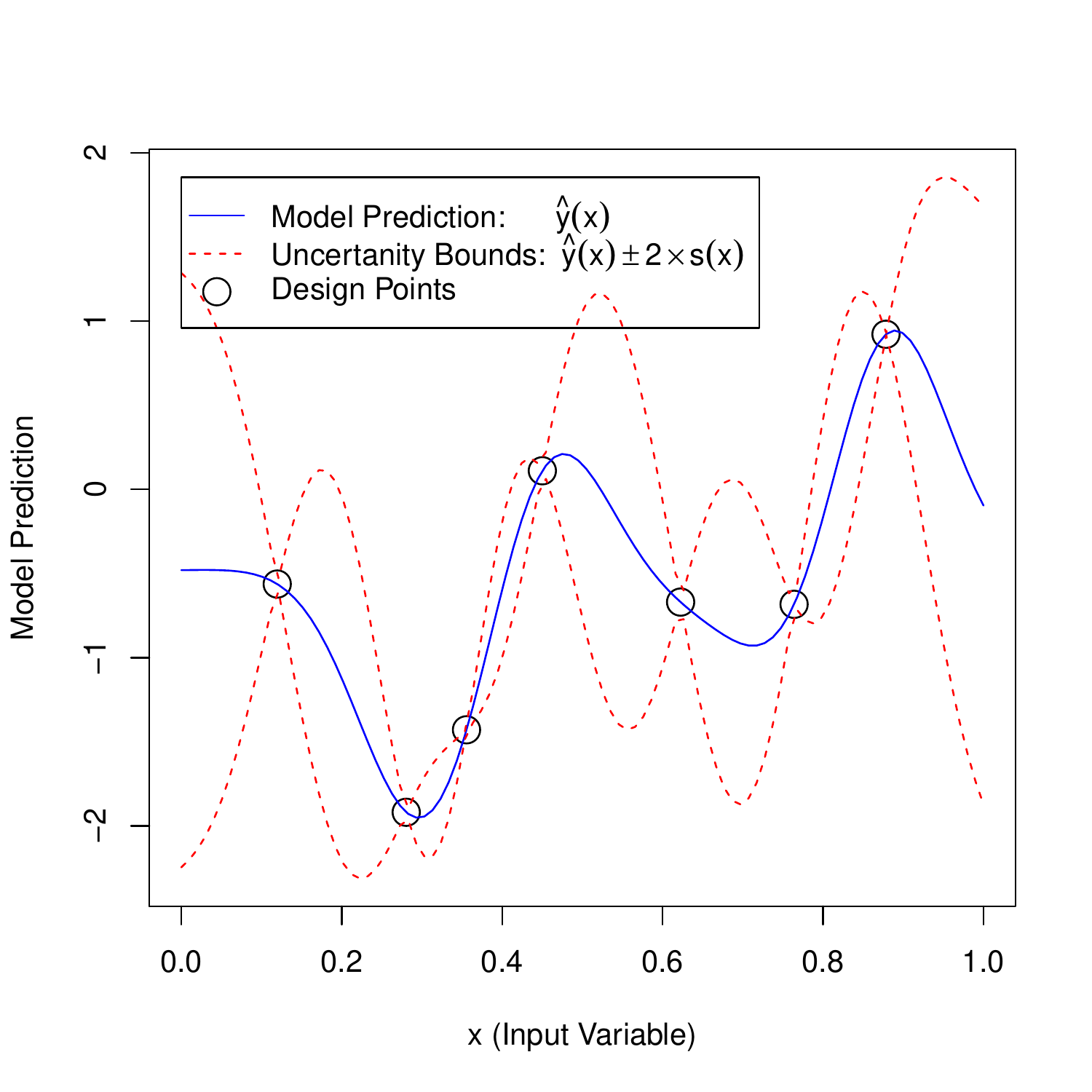}}\vspace{-0.5cm}
\subfigure[\code{cex=3}]{\label{fig:eg1c} \includegraphics[width=3in]{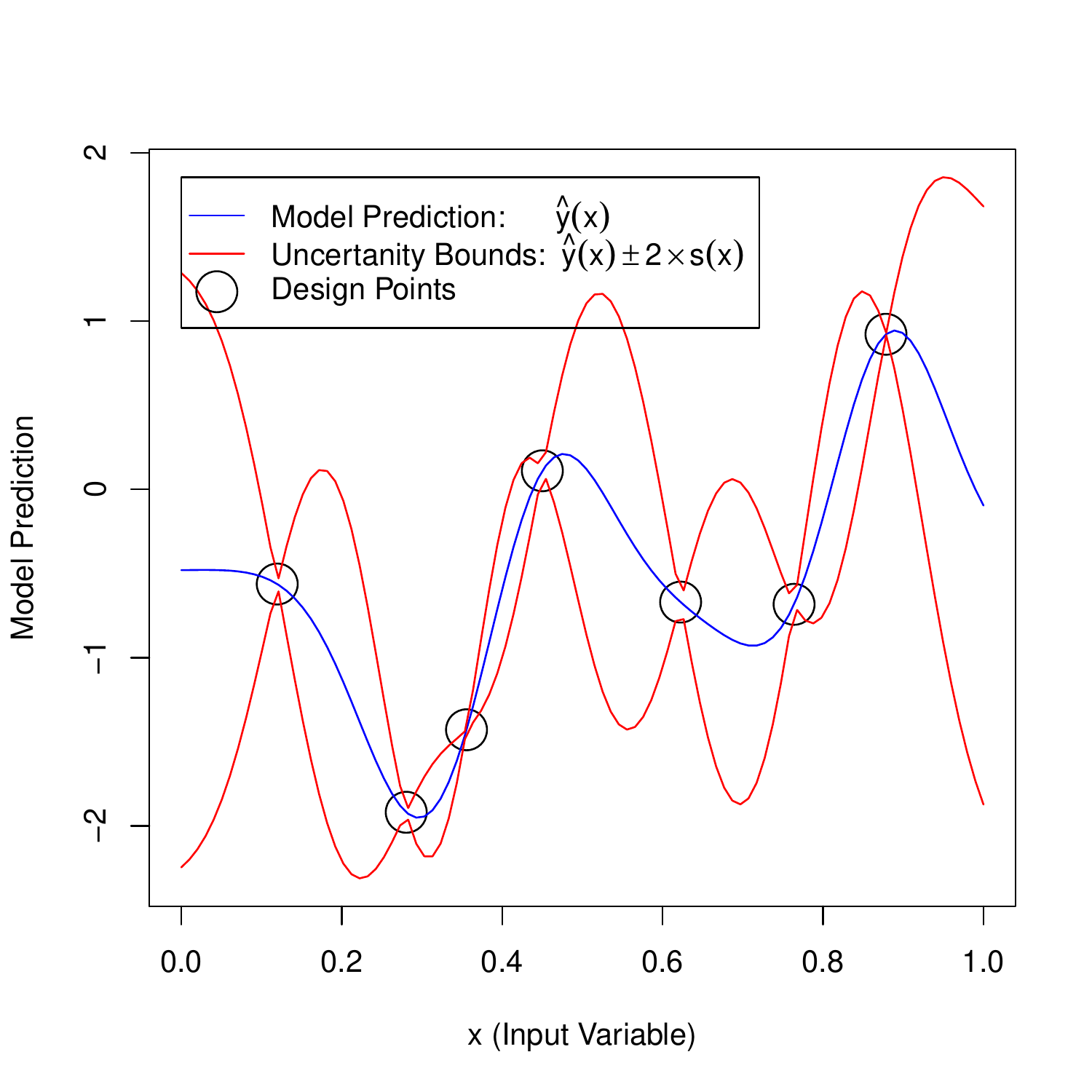}}\hspace{-0.75cm}
\subfigure[\code{line\_type=c(1,2), pch=2, cex=3}]{\label{fig:eg1d} \includegraphics[width=3in]{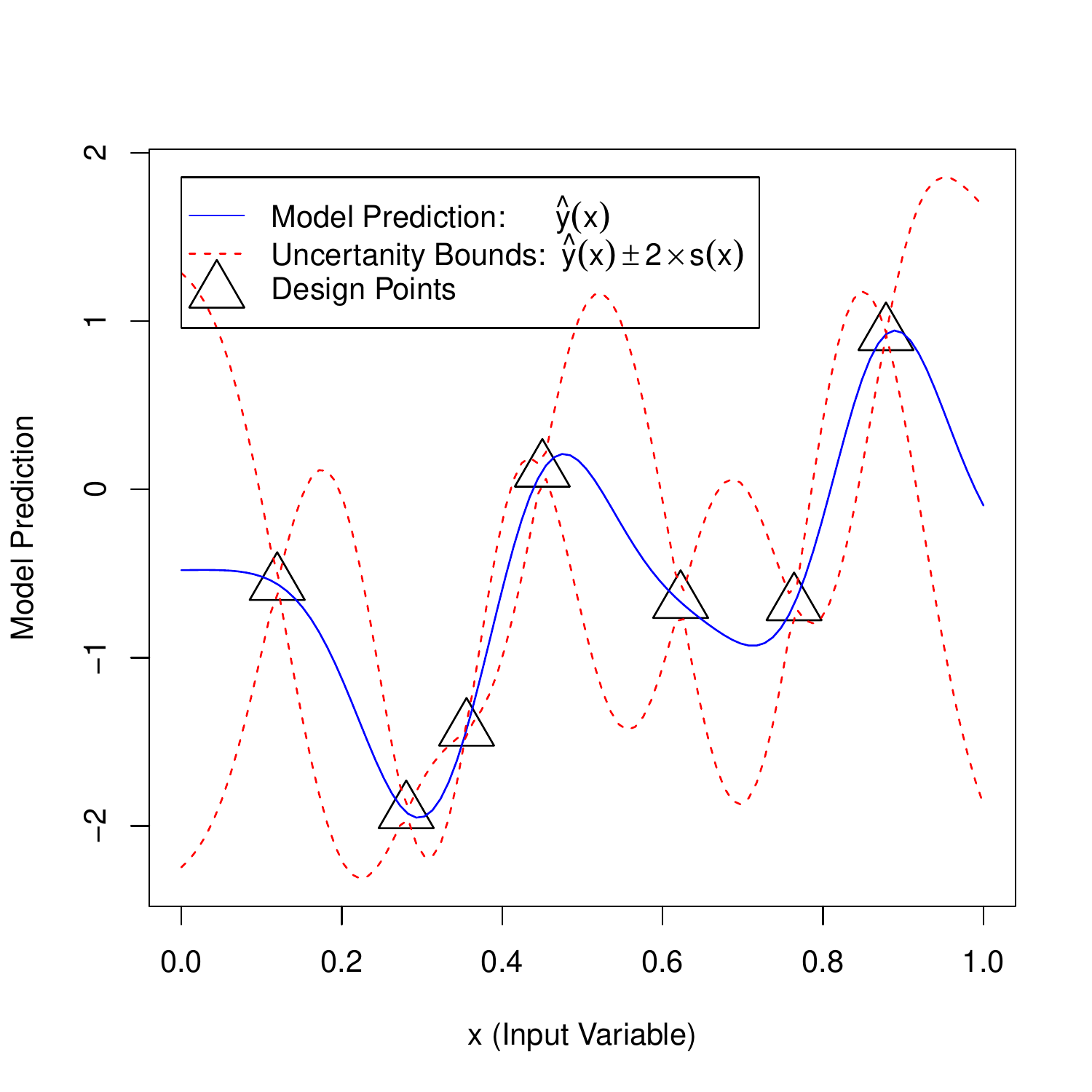}}
\caption{The plots illustrate the usage of graphical parameters in \code{plot()} for Example~1. Panel (a) shows the model prediction and uncertainty plot with default graphical parameters, (b) illustrates the change due to \code{line\_type}, (c) highlights the point size using \code{cex}, and (d) shows the usage of \code{pch} in changing the point character.}\label{fig:eg1}\end{figure}

The \verb"GPmodel" object can be used to predict and then plot the simulator outputs at a grid of inputs using \verb"`plot(GPmodel,...)'". Figures~\ref{fig:eg1_0} and \ref{fig:eg1} show the model prediction along with the uncertainty bounds $\hat{y}(x^*) \pm 2s(x^*)$ on the uniform grid with \verb"`resolution=100'". Figure~\ref{fig:eg1_0} compares the predicted and the true simulator output. Figure~\ref{fig:eg1} illustrates the usage of the graphical arguments of \verb"plot()". \verb"`predict(GPmodel,xnew)'" can also be used to obtain model predictions at an arbitrary set of inputs, \verb"xnew", in the design space (i.e., not a grid).

%

\textbf{Example 2} We now consider a two-dimensional test function to illustrate different functions of \pkg{GPfit} package. Let $x=(x_1, x_2) \in [-2, 2]^2$, and the simulator outputs be generated from the GoldPrice function \citep{andre}
\begin{eqnarray*}
&y(x)& = \left[ 1 + \left({x_1} + {x_2} + 1 \right)^2\left\{19-{14x_1} + 3x_1^2- 14x_2 + 6x_1x_2+3x_2^2\right\}\right]* \\
&&\left[30+\left(2x_1 - 3x_2\right)^2(18 - 32x_1 + 12x_1^2 + 48x_2 - 36x_1x_2+27x_2^2)\right].
\end{eqnarray*}
For convenience the inputs are scaled to $[0, 1]^2$. The \verb"GP_fit()" output from fitting the GP model to a data set based on a $20$-point maximin LHD is as follows:
\begin{CodeChunk}
\begin{CodeOutput}
Number Of Observations: n = 20
Input Dimensions: d = 2

Correlation: Exponential (power = 2)
Correlation Parameters:
    beta_hat.1 beta_hat.2
[1]     0.8578      1.442

sigma^2_hat: [1] 4.52e+09
delta_lb(beta_hat): [1] 0
nugget threshold parameter: 20
\end{CodeOutput}
\end{CodeChunk}

For fitting this GP model, the proposed multi-start L-BFGS-B optimization procedure used only 808 deviance evaluations, whereas the GA based optimization in \cite{ranjanNugget} would have required $4000$ deviance calls. The correlation hyper-parameter estimate $\hat{\beta}_{mle} = (0.8578, 1.442)$ shows that the fitted simulator is slightly more active (or wiggly) in the $X_2$ variable. The nugget parameter $\delta_{lb}(\hat{\beta}_{mle}) = 0$ implies that the correlation matrix with the chosen design points and $\beta = \hat{\beta}_{mle}$ is well-behaved.

The following code illustrates the usage of \verb"predict()" for obtaining predicted response and associated MSEs at a set of unobserved inputs.
\begin{CodeChunk}
\begin{CodeInput}
R> xnew = matrix(runif(20),ncol=2)
R> Model_pred = predict(GPmodel,xnew)
\end{CodeInput}
\end{CodeChunk}
The model prediction outputs stored in \verb"predict" object \verb"Model_pred" are as follows:
\begin{CodeChunk}
\begin{CodeOutput}
$Y_hat
 [1]    561.3877   -372.5221  13287.0495   3148.5904   5129.1136
 [6]   8188.2805   3626.4985  14925.8142   2869.6225 217039.3229

$MSE
 [1]  186119713   21523832   86391757    8022989  562589770
 [6]   13698589  123121468 1167409027 1483924477  264176788

$complete_data
         xnew.1    xnew.2       Y_hat        MSE
 [1,] 0.2002145 0.2732849    561.3877  186119713
 [2,] 0.6852186 0.4905132   -372.5221   21523832
 [3,] 0.9168758 0.3184040  13287.0495   86391757
 [4,] 0.2843995 0.5591728   3148.5904    8022989
 [5,] 0.1046501 0.2625931   5129.1136  562589770
 [6,] 0.7010575 0.2018752   8188.2805   13698589
 [7,] 0.5279600 0.3875257   3626.4985  123121468
 [8,] 0.8079352 0.8878698  14925.8142 1167409027
 [9,] 0.9565001 0.5549226   2869.6225 1483924477
[10,] 0.1104530 0.8421794 217039.3229  264176788
\end{CodeOutput}
\end{CodeChunk}

The \pkg{GPfit} function \verb"plot()" calls \verb"predict()" for computing $\hat{y}(x^*)$ and $s^2(x^*)$ at a regular \verb"`resolution x resolution'" grid in the input space defined by the \verb"`range'" parameter. Recall from Section~\ref{sec:gpfit} that \verb"colors", \verb"line_type", \verb"pch" and \verb"cex" are only applicable for  one dimensional plots. For $d=2$, the following code can be used to draw the level/contour and surface plots of $\hat{y}(x)$ and $s^2(x)$ over a specified grid resolution.
\begin{CodeChunk}
\begin{CodeInput}
    plot(GPmodel, range=c(0,1), resolution=50, surf_check=FALSE,
                                                    response=TRUE, ...)
\end{CodeInput}
\end{CodeChunk}

\begin{figure}[htb!]\centering
\subfigure[Default arguments]{\label{fig:eg2a} \includegraphics[width=3in]{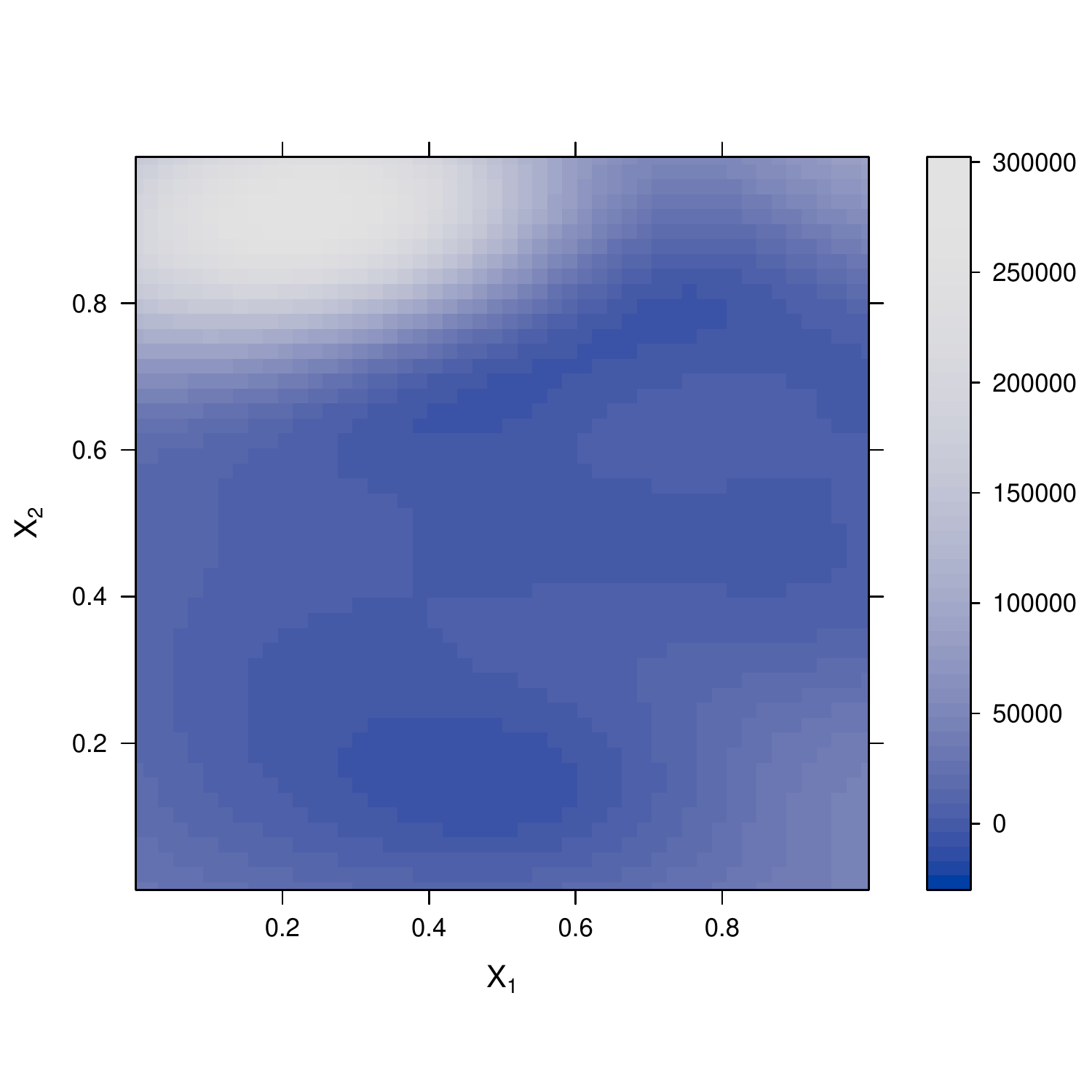}}\hspace{-0.3cm}
\subfigure[\code{response=FALSE, contour=TRUE}]{\label{fig:eg2b} \includegraphics[width=3in]{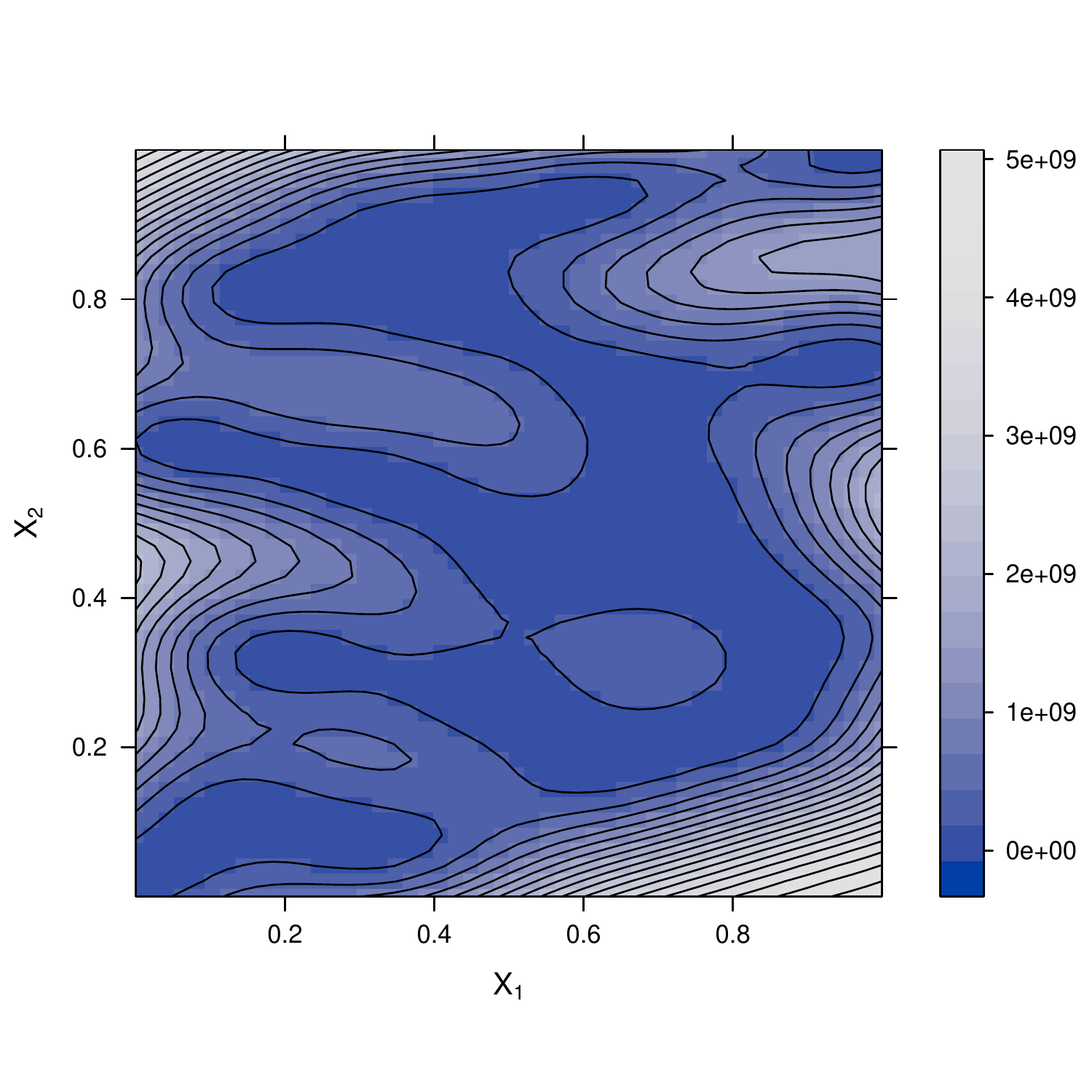}}\vspace{-0.25cm}
\subfigure[\code{surf\_check=TRUE}]{\label{fig:eg2c} \includegraphics[width=3in]{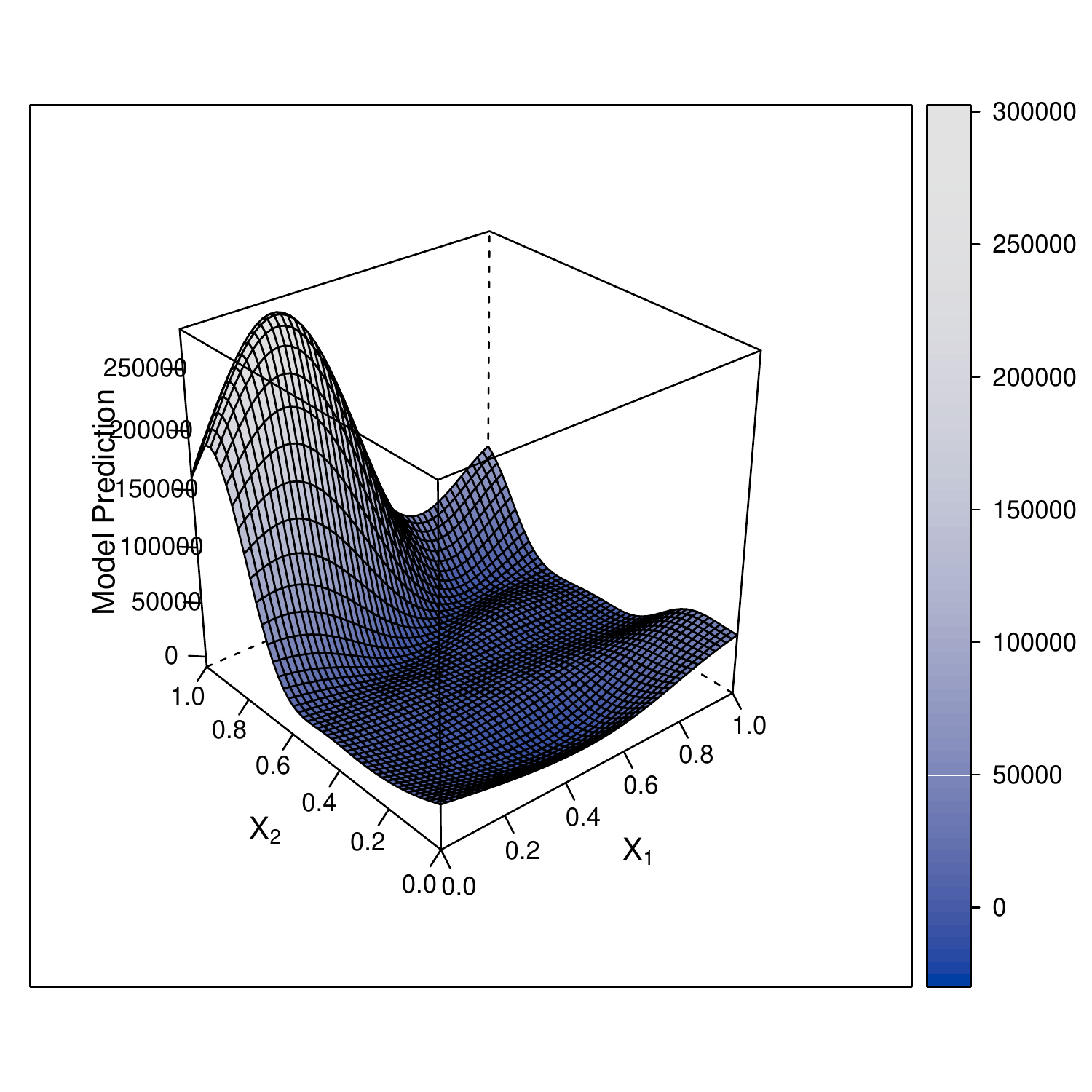}}\hspace{-0.5cm}
\subfigure[\code{response=FALSE, surf\_check=TRUE}]{\label{fig:eg2d} \includegraphics[width=3in]{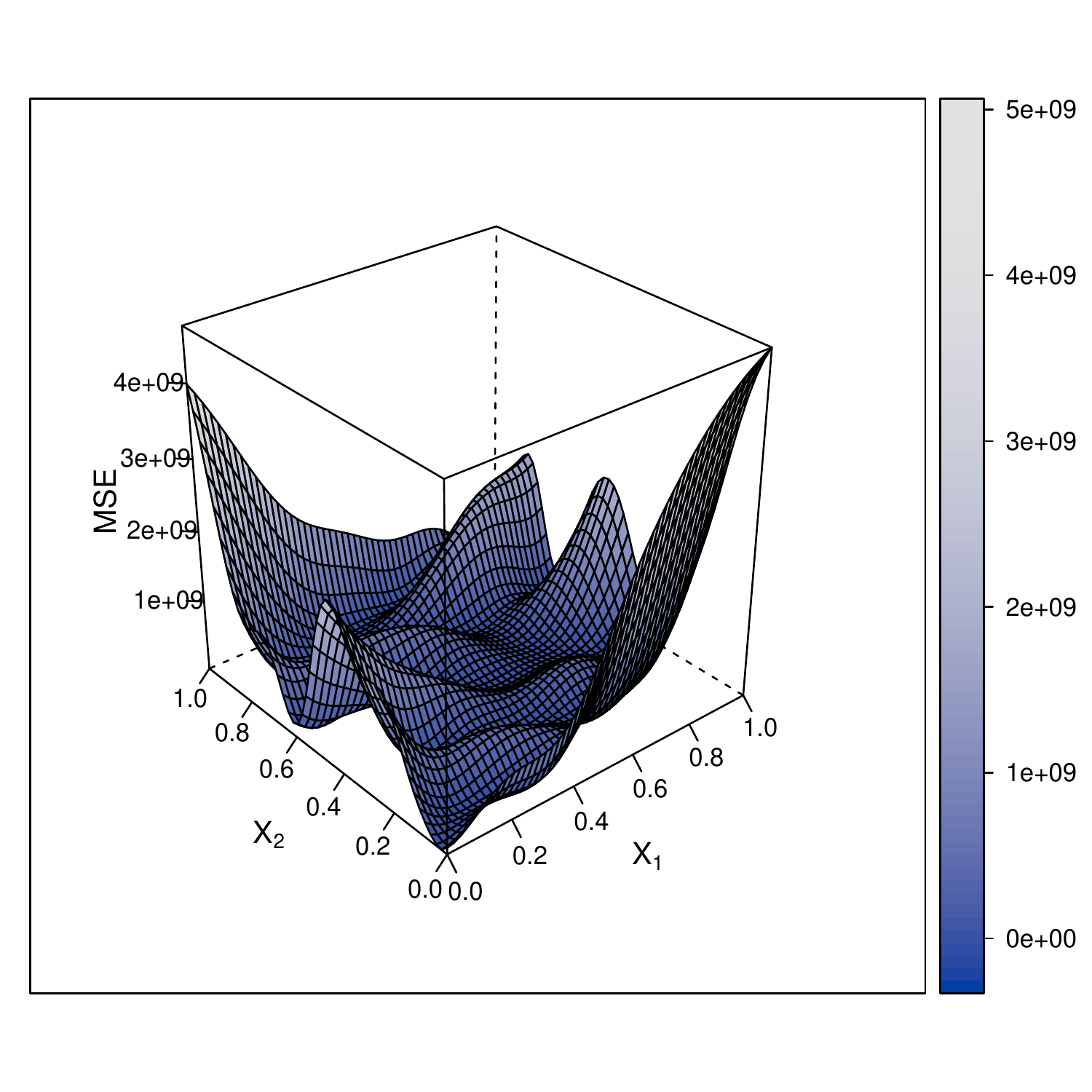}}
\caption{The plots illustrate the usage of graphical parameters in \code{plot()} for Example~2. Panel (a) shows the default plot (the levelplot of $\hat{y}(x^*)$) with additional color specification, (b) presents levelplot with contour lines of $s^2(x^*)$, (c) shows the surface plot of $\hat{y}(x^*)$, and (d) displays the surface plot of $s^2(x^*)$.}\label{fig:eg2}\end{figure}

Additional graphical arguments, for instance, from \pkg{lattice} and \pkg{colorspace}, can also be passed in for ``..." to enhance the plotting features. Figure~\ref{fig:eg2} shows the model predictions and the MSEs on the uniform $50 \times 50$ grid. Figures~\ref{fig:eg2a} and \ref{fig:eg2b} used additional argument \code{`col.regions=sequential\_hcl(51, power=2.2)'} (from \pkg{colorspace} package) to change the default color palettes. Different panels of Figure~\ref{fig:eg2} highlight the usage of \code{surf\_check} and \code{response} for obtaining a level plot and surface plot of $\hat{y}(x)$ and $s^2(x)$.

%


\section{Comparison with other packages}\label{sec:mlegp_comp}
In the last two decades, a few different programs (in \proglang{R,
Matlab, C, C++, Python}, and so on) have been produced for fitting
GP models in computer experiments. The Gaussian process website
\citep{GPwebsite} presents an extensive (though incomplete) list of
such programs. Since \proglang{R} is a free software environment, packages like \pkg{tgp} and \pkg{mlegp} have gained popularity among
the practitioners in computer experiments.

The \pkg{tgp} package \citep{tgp_paper,gramacy}, originally developed for building surrogates of both stationary and non-stationary stochastic (noisy) simulators, uses a GP model for emulating the stationary components of the process. The GP model here includes a nugget parameter that is estimated along with other parameters. The recent version of the \pkg{tgp} package facilitates the emulation of deterministic simulators by removing the nugget parameter from the model. Most importantly, \pkg{tgp} is implemented using Bayesian techniques like Metropolis-Hastings algorithm, whereas, \pkg{GPfit} follows the maximum likelihood approach for fitting GP models and includes the smallest possible nugget required for computational stability.

\cite{mlegp} developed an \proglang{R} package called \pkg{mlegp} that uses
maximum likelihood for fitting the GP model with Gaussian correlation
structure. Though not relevant for this paper, \pkg{mlegp} can fit
GP models with multivariate response, non-constant mean function
and non-constant variance that can be specified exactly or up to a
multiplicative constant.  The simple GP model in \pkg{mlegp} is the
same as described in Section~\ref{sec:GPmodel} except that the nugget
parameter is estimated along with other hyper-parameters. Hence, we use
\pkg{mlegp} for the performance comparison of \pkg{GPfit}.

We now use several test functions to compare the performance of the two packages \pkg{mlegp} and \pkg{GPfit}. The test functions used here are commonly used in computer experiments for comparing competing methodologies \citep{santner2003daa}. Since the two packages minimize slightly different deviance functions, one cannot directly compare the parameter estimates or the minimized deviance. Consequently, we compared the discrepancy between the predicted and the true simulator response. The performance measure is the standardized/scaled root mean squared error (sRMSE) given by
$$
\frac{1}{y_{max} - y_{min}} \sqrt{\frac{1}{N}\sum_{i=1}^{N} \left[\hat{y}(x_i^*) - y(x_i^*)\right]^2},
$$
where $y_{max}$ and $y_{min}$ are the global maximum and minimum of the
true simulator, $y(x_i^*)$ and $\hat{y}(x_i^*)$ are the true and predicted
simulator output at $x_i^*$ in the test data, and $N$ is the size of
the test data set. The results are averaged over 50 simulations. Each
simulation starts with choosing two random $n \times d$ maximin LHDs
($D_0$ and $D_1$) for the training data and test data respectively
(i.e., $N=n$). The average and standard error of the sRMSE values of
the GP fits obtained from \pkg{mlegp} and \pkg{GPfit} are compared
for several design sizes.

We found that \pkg{mlegp} occasionally crashes due to near-singularity of the spatial correlation matrix in the GP model. In \pkg{mlegp}, the nugget parameter in $R_{\delta} = R + \delta I$ is estimated using maximum likelihood procedure along with the other model parameters. If any candidate $\delta \in (0, 1)$ in the optimization procedure is not large enough to overcome the ill-conditioning of $R_{\delta}$, the likelihood computation fails and the \pkg{mlegp} package crashes with the following error message:

\begin{verbatim}
Error in solve.default(gp\$invVarMatrix):
system is computationally singular:~reciprocal condition number = 2.11e-16.
\end{verbatim}

This is not a problem in \pkg{GPfit} implementation, because the nugget parameter is set at the smallest $\delta$ required to make $R_{\delta}$ well-conditioned. As a result, \pkg{GPfit} outperforms \pkg{mlegp} in terms of computational stability. Whenever \pkg{mlegp} runs are computationally stable, then also \pkg{GPfit} appears to have lower sRMSE values in most cases.\\

\textbf{Example 1 (contd.)} Suppose we wish to compare the prediction accuracy of the GP model fits from the two packages for the one dimensional test function in Example~1. Table~\ref{tab:example3} summarizes the sRMSE values for a range of sample sizes in the format: average (standard error). The results are based on $50$ simulations.

\begin{table}[!ht]\centering
\begin{tabular}{|l|r|r|}
\hline
\multirow{3}{*}{Sample size} & \multicolumn{1}{|c|}{GPfit} & \multicolumn{1}{|c|}{mlegp} \\
\cline{2-3}
& {sRMSE ($\times 10^{-6}$)}  & \multicolumn{1}{|c|}{sRMSE ($\times 10^{-6}$)} \\
\hline
$n$ = 10 & 32958 (4948.8) & 37282 (7153.4)       \\
\hline
$n$ = 25 & 139.21 (13.768)  &   158.07 (15.662)      \\
\hline
$n$ = 50 & 28.81 (2.5977)   &  113.49 (16.139)     \\
\hline
$n$ = 75 & 18.29 (1.6297)    & 105.84 (16.251)      \\
\hline
$n$ = 100 & 12.36 (0.7320)    &  101.25 (14.254)    \\
\hline
\end{tabular}\caption{The summary of sRMSE values for the one dimensional simulator in Example~1.}\label{tab:example3}
\end{table}

It is clear from Table~\ref{tab:example3} that the sRMSE values decrease in both methods as $n$ increases. More importantly, \pkg{GPfit} significantly outperforms \pkg{mlegp}, especially, for larger $n$. This is expected as the numerical instability of the GP model increases with $n$. The smallest nugget $\delta_{lb}$ in the GP model of \pkg{GPfit} minimizes unnecessary over-smoothing hence smaller sRMSE as compared to that in \pkg{mlegp}, where $\hat{\delta}_{mle}$ might be relatively large to ensure computationally stable GP model fits (i.e., without any crashes).\\

\textbf{Example 2 (contd.)} We now revisit the two-dimensional GoldPrice function illustrated in Example~2. Table~\ref{tab:example4} presents the averages and standard errors of sRMSE values for GP model fits obtained from \pkg{mlegp} and \pkg{GPfit}.

\begin{table}[!ht]\centering
\begin{tabular}{|l|r|r|r|}
\hline
\multirow{3}{*}{Sample size} & \multicolumn{1}{|c|}{GPfit} & \multicolumn{2}{|c|}{mlegp} \\
\cline{2-4}
& {sRMSE ($\times 10^{-4}$)} & \multicolumn{1}{|c|}{sRMSE ($\times 10^{-4}$)} & Crashes\\
\hline
$n$ = 25 &381.23 (43.85)    &     424.07 (56.92) &       0\\
\hline
$n$ = 50&88.120 (8.114)   &  105.95 (18.93)   &     0\\
\hline
$n$ = 75 &23.282 (1.499) &    17.379 (2.271)      &  0\\
\hline
$n$ = 100 &12.747 (0.875)&        1601.5 (188.6)       &14\\
\hline
\end{tabular}\caption{The summary of sRMSE values and the number of crashes for GoldPrice function.}\label{tab:example4}
\end{table}

It is important to note that the \pkg{mlegp} crashed 14 times out of 50 simulations for the $n=100$ case. The summary statistics for $n=100$ case in the \pkg{mlegp} column are calculated from the remaining 26 successful runs. The average and standard error of the sRMSE values in the successful runs of \pkg{mlegp} generate unreliable predictions. For the remaining cases, the results show that the sRMSE values decrease in both methods as $n$ increases. For $n=25$ and $50$, \pkg{GPfit} produces better GP fits with smaller sRMSE values. Interestingly, for $n=75$, the average sRMSE value in \pkg{GPfit} is slightly larger as compared to that in \pkg{mlegp}. \\
%

\textbf{Example 3} Suppose the four-dimensional Colville function is used as the computer simulator. Let $x=(x_1, x_2,x_3,x_4) \in [-10, 10]^4$, and the outputs be generated from
\begin{eqnarray*}
y(x) &=&  100(x_1^2 - x_2)^2 + (x_1 - 1)^2 +(x_3 - 1)^2+90(x_3^2-x_4)^2\\
&& + 10.1[(x_2-1)^2+(x_4-1)^2] +19.8(x_4-1)/x_2.
\end{eqnarray*}

For implementation purpose, the inputs are rescaled to the unit-hypercube $[0, 1]^4$. Table~\ref{tab:example5} summarizes the averages and standard errors of the sRMSE values from 50 simulations.

\begin{table}[!ht]\centering
\begin{tabular}{|l|r|r|r|}
\hline
\multirow{2}{*}{Sample size} & \multicolumn{1}{|c|}{GPfit} & \multicolumn{2}{|c|}{mlegp} \\
\cline{2-4}
& \multicolumn{1}{|c|}{sRMSE ($\times 10^{-6}$)} &\multicolumn{1}{|c|}{sRMSE ($\times 10^{-6}$)} & Crashes\\
\hline
$n$ = 25 &103.3 (5.401) &        109.58 (6.120)&       0\\
\hline
$n$ = 50 & 11.77 (0.771)&        10334 (3344)&        2\\
\hline
$n$ = 75 &  7.169 (0.472)&        3251 (1109)&        5\\
\hline
$n$ = 100 &5.786 (1.839)&        63.10 (25.39)&        1\\
\hline
\end{tabular}\caption{The summary of sRMSE values and the number of crashes for Colville function.}\label{tab:example5}
\end{table}

Similar to Example~2, a few runs from \pkg{mlegp} crashed
due to near-singularity, and the successful runs in these cases ($n=50,
75$ and $100$) yield unreliable summary statistics (i.e., unrealistically
large sRMSE values).  In contrast, \pkg{GPfit}
provides stable and good predictions. Similar to Examples~1
and 2, the average sRMSE values decrease as $n$ increases.

It is worth noting that for the $n=100$ case in this example, \pkg{mlegp} crashed only once in 50 simulations, whereas for the GoldPrice function example (Table~\ref{tab:example4}), \pkg{mlegp} crashed 14 times. Though the number of simulations considered here is not large enough to accurately estimate the proportion of crashes in each case, it is expected that the occurrence of near-singular cases becomes less frequent with the increase in the input dimension (see \cite{ranjanNugget} for more details).\\


\textbf{Example 4} Consider the six-dimensional Hartmann function for generating simulator outputs. Since the input dimension is reasonably large, all \pkg{mlegp} runs turned out to be successful, and both the packages lead to similar model predictions. Table~\ref{tab:example6} presents the averages and standard errors of the sRMSE values.

\begin{table}[!ht]\centering
\begin{tabular}{|l|r|r|}
\hline
\multirow{2}{*}{Sample size} & \multicolumn{1}{|c|}{GPfit} & \multicolumn{1}{|c|}{mlegp} \\
\cline{2-3}
& \multicolumn{1}{|c|}{sRMSE ($\times 10^{-3}$)}  &  \multicolumn{1}{|c|}{sRMSE ($\times 10^{-3}$)}\\
\hline
$n$ = 25 &118.44 (4.837)&       116.64 (4.655)\\
\hline
$n$ = 50 &105.24 (4.649)&       105.56 (4.500)\\
\hline
$n$ = 75 &82.587 (2.536)&       84.819 (3.090)\\
\hline
$n$ = 100 & 75.169 (2.645)&       75.402 (2.738)\\
\hline
$n$ = 125 &63.014 (1.652)&       63.223 (1.653)\\
\hline
\end{tabular}
\caption{The summary of sRMSE values for the six-dimensional Hartmann function.}\label{tab:example6}
\end{table}


Overall in Examples~1 to 4, \pkg{mlegp} crashed only 22 times out of 900 simulations. However, the successful runs in the cases with any crash ($n=100$ in Example~2 and $n = 50, 75$ and $100$ in Example~3) lead to unreliable model fits. Furthermore, \pkg{GPfit} either outperforms or gives comparable GP model fits as compared to \pkg{mlegp}.

\section{Concluding remarks}\label{sec:conclusion}

This paper presents a new \proglang{R} package \pkg{GPfit} for fitting GP models to
scalar valued deterministic simulators. \pkg{GPfit} implements a slightly
modified version of the GP model proposed by \cite{ranjanNugget}, which
uses the new $\beta$ parameterization (\ref{eqn:corr_beta}) of the spatial correlation function for the ease of optimization. The deviance optimization is achieved through a multi-start L-BFGS-B algorithm.

The proposed optimization algorithms makes $200d + \sum_{i=1}^{2d+1}\eta_i
+ \sum_{j=1}^{3}\eta_j'$ calls of the deviance function, whereas the GA
implemented by \cite{ranjanNugget} uses $1000d^2$ deviance evaluations.
Though $\eta_i$ and $\eta_j'$ are  non-deterministic, and vary with the
complexity and input dimension of the deviance surface, the simulations in
Section~\ref{sec:mlegp_comp} show that $\eta_j' \approx 30$ for all
examples, however, the average $\eta_i$ are approximately $40, 75, 300$ and
interestingly $150$ for Examples~1, 2,
3 and 4 respectively. Of course, neither of the
two implementations have been optimally tuned for the most efficient
deviance optimization.
The best choice of options will of course vary from problem to
problem, and so we encourage users to experiment with the available
options.

The \pkg{mlegp} package is written in pre-compiled \proglang{C} code, whereas \pkg{GPfit} is implemented solely in \proglang{R}. This difference in the programming environment makes \pkg{mlegp} substantially faster than \pkg{GPfit}. The current version of \pkg{GPfit} package uses only Gaussian correlation. We intend to include other popular correlation functions like Mat\'ern in our \proglang{R} package.

\bibliographystyle{jss}
\bibliography{hello}

\end{document}